\documentclass[APS,twocolumn,showpacs,preprintnumbers]{revtex4}
\pdfoutput=1
\usepackage{graphicx}
\usepackage{psfrag}
\usepackage{amsfonts,amssymb, amsmath}
\usepackage{amsthm}

\newcommand{\id}{\openone}
\newcommand{\bra}[1]{\langle #1|\,}
\newcommand{\ket}[1]{\,|#1 \rangle}

\newcommand{\ketbra}[2]{\,|#1 \rangle\negmedspace\langle #2|\,}

\newcommand{\defeq}{\stackrel{\text{def}}{=}}

\newcommand{\avcon}[1]{\bar{C}_{#1}(x)}
\newcommand{\Cf}{\mathbb{C}^4}

\newcommand{\Cqm}{C^{\text{QM}}}
\newcommand{\Ccl}{C^{\text{Direct}}}
\newcommand{\Cspp}[1]{C^{\text{SPP}}_{#1}}

\newcommand{\yhi}{ {y^*_{\text{h}}}}
\newcommand{\ylo}{ {y^*_{\text{l}}}}
\newcommand{\spath}{{\cal{P}}_{AB}}

\newcommand{\grayfig}{_gray}  % use this one for gray figures

\DeclareMathOperator{\tr}{tr}

\begin{document}

\title{Distribution of entanglement in networks of bi-partite full-rank mixed states}
\author{G.J. Lapeyre, Jr.$^1$, S. Perseguers$^2$, M. Lewenstein$^{1,3}$, A. Ac\'in$^{1,3}$}
\affiliation{
    $^1$ICFO--Institut de Ci\`encies Fot\`oniques, Mediterranean Technology Park, 08860 Castelldefels, Spain\\
    $^2$Max-Planck--Institut f\"ur Quantenoptik, Hans-Kopfermann-Strasse 1, 85748 Garching, Germany\\
    $^3$ICREA-Instituci\'o Catalana de Recerca i Estudis Avan\c cats, Lluis Companys 23, 08010 Barcelona, Spain
}
\date{\today}

% =============================================================================
% ABSTRACT
% =============================================================================
\begin{abstract}
  We study quantum entanglement distribution on networks
  with full-rank bi-partite mixed states linking qubits on
  nodes. In particular, we use entanglement swapping and
  purification to partially entangle widely separated
  nodes. The simplest method consists of performing entanglement swappings
  along the shortest chain of links connecting the two nodes.
  However, we show that this method may be improved upon by
  choosing a protocol with a specific ordering  of swappings and purifications.
  A priori, the design that produces optimal improvement is not clear.
  However, we parameterize the choices and find that the optimal
  values depend strongly on the desired measure of improvement. As an initial application,
  we apply the new improved protocols to the Erd\"os--R\'enyi network and obtain results including
  low density limits and an exact calculation of the average entanglement gained at the
  critical point.
\end{abstract}

\maketitle

% =============================================================================
% MAIN TEXT
% =============================================================================

% Introduction
% -----------------------------------------------------------------------------

\section{Introduction}

The quantum repeater has been at the center of numerous studies
addressing the distribution of entanglement over long distances,
an essential prerequisite for many tasks in quantum information
processing~\cite{DBCZ99,BDCZ98}. The central idea of the quantum
repeater is to send one of a pair of entangled particles ({\it
e.g.} a photon) across a series of links such that each link is
short enough that the probability of absorption is low and then to
perform entanglement swappings at each node to further propagate
the entanglement.  However, the inevitable presence of noise in
producing and transporting quantum states renders the
straightforward application of repeaters hard in practice. This
has resulted in the expenditure of a great deal of effort in
designing distribution protocols.  While the initial repeater
schemes involved a one-dimensional chain of nodes (containing
qubits) connected by links, considering higher-dimensional
networks of nodes and links has been a fruitful approach because
the entanglement in neighboring links can be concentrated via
purification. Given a fixed amount of entanglement per link that
can be generated in a particular laboratory setting, this
concentration allows a useful amount of entanglement to be
distributed over larger distances, which in turn allows protocols
that consume entanglement to work over a larger distance. Previous
work on entanglement distribution in dimension greater than one
considered either pure
states~\cite{Natphys.3.1745,perseguers:022308,lapeyre_enhancement_2009},
or certain mixed states of rank two and three~\cite{BDJ09,BDJ10}.
Pure states were used in previous distribution studies because
entanglement is better understood and easier to manipulate in pure
states than mixed states.  However, realistic noise models imply
that a bi-partite system of non-local components may only be
prepared in a full-rank mixed state.  In fact, it has been shown
that long-range entanglement on a cubic lattice of full-rank mixed
states is possible~\cite{Perseguers2010a}.

On the other hand, while the bulk of work to date has been concerned with
regular lattices, only a few works have treated random networks. The creation of
entangled sub-networks on the Erd\"os--R\'enyi (ER)
model~\cite{perseguers_quantum_2010} has been studied.
Other workers investigated the effect of a particular transformation on several pure--state
complex networks as well entanglement swapping with full-rank
mixed states on the same networks~\cite{PhysRevLett.103.240503,CC11}. % cuquet, calsamigili
These studies showed on one hand that transforming network hubs to rings via LOCC  can enhance
entanglement distribution on a variety of pure-state networks and on the
other hand how distribution is affected by the interplay between the correlation
length and the characteristic length of decay of fidelity under swapping.
However, detailed studies of the application of of entanglement distribution
protocols using both swapping and purification on
complex networks have not yet been done.  In the present work we
address this deficit by introducing some natural optimization
problems in distributing entanglement along paths of full-rank mixed states. We
find that the solutions to these problems yield surprisingly non-trivial
results. Then, as a first application, we apply entanglement concentration
protocols to the ER network. One of our main objectives is to understand when
the use of the network connectivity offers an advantage for distributing
entanglement between two nodes with respect to the simple protocol in which
entanglement is swapped along the shortest path connecting them.  Note that this
is always the case for a classical network: connectivity always helps in
distributing classical information through a network. However, this may not be
the case in the quantum regime, as quantum information cannot be
cloned~\cite{nocloning}. Indeed, we provide instances where the simplest direct
protocol is better than the considered protocols using the network connectivity.

\begin{figure}
    \begin{center}
        \includegraphics[width=0.5\linewidth]{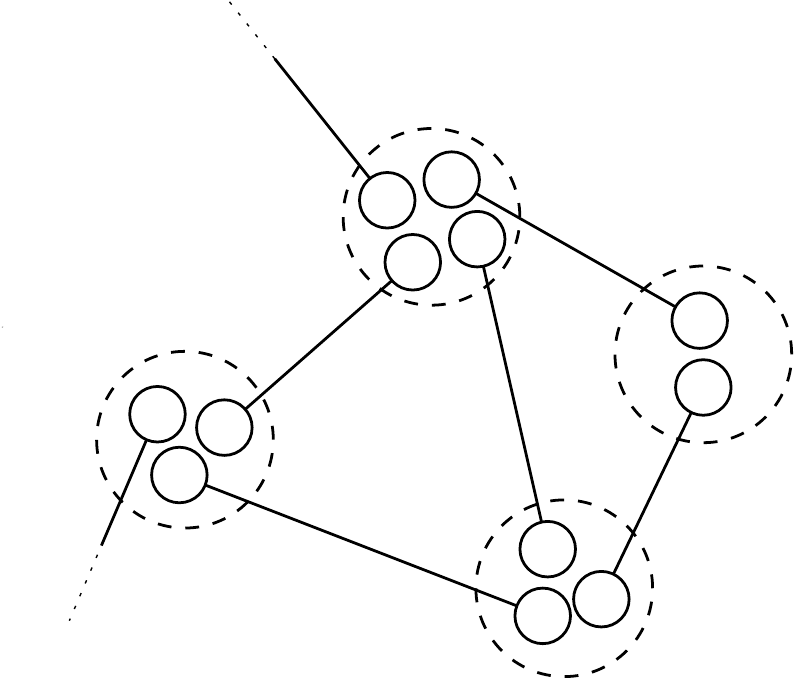}
        \caption{ Part of a quantum network in its initial
          state. Small circles are qubits. Solid lines are
          bipartite states. Dashed circles enclose qubits
          within a node. Local operations may act on all
          qubits within a node.  }
    \label{fig:generic_network}
    \end{center}
\end{figure}
In Sec.~\ref{sec:elements}, we give an overview of the models that we will examine.
We shall consider the average  concurrence on
networks in which each link is initially a full-rank mixed state on
two two-level systems, while a node is a local collection consisting
of one party from each link terminating at that node. (See Fig.~\ref{fig:generic_network}.) Unless stated
explicitly, when we speak
of average concurrence we mean an average over both the outcomes of quantum
measurements and the distribution of links for a given random network.
\begin{figure}
    \begin{center}
        \includegraphics[width=0.75\linewidth]{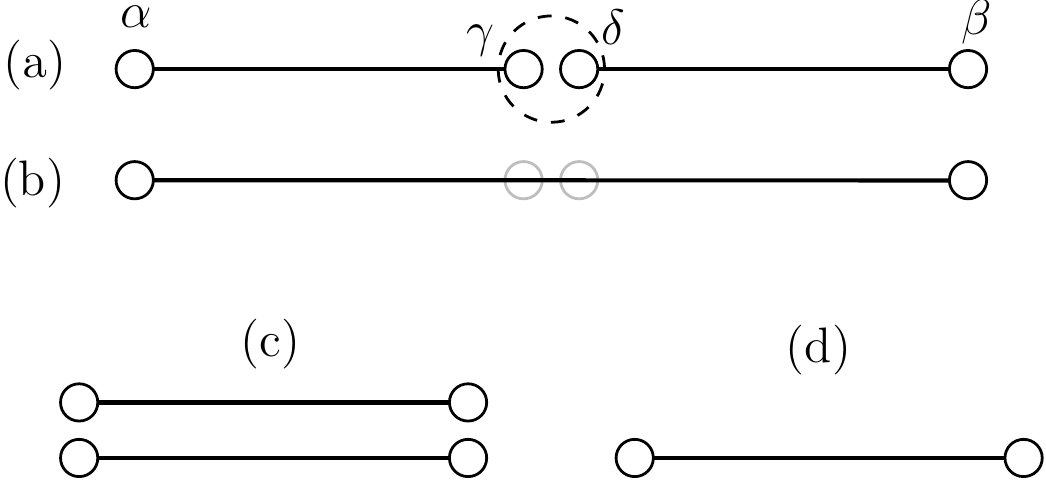}
        \caption{ Entanglement swapping: (a) Before swapping
          $\alpha$ and $\gamma$ are entangled and $\delta$
          and $\beta$ are entangled, but systems
          $\alpha\gamma$ and $\delta\beta$ are in a product
          state.  (b) After swapping, systems $\alpha$ and
          $\beta$ are entangled, while $\alpha\beta$ and
          $\gamma\delta$ are in a product state.
          Purification: (c) Before purification, two
          entangled pairs (links). (d) after purification,
          one pair of nodes has higher entanglement than
          either of the original pairs.}
    \label{fig:swappure}
    \end{center}
\end{figure}
We shall furthermore consider only two quantum operations for
distributing entanglement (See Fig. \ref{fig:swappure}.):
\begin{itemize}
 \item entanglement swapping, which
probabilistically replaces a series of two links by a single link
that bypasses the common node. The output link is in general
 less entangled than the input links.
 \item  purification, which
essentially replaces two parallel links ({\it i.e.} sharing the same two nodes) by a single link that is more
highly entangled than either input link.
\end{itemize}
The main reason for the restriction to two operations is that many techniques
that are successful on pure states, such as multi-partite techniques~\cite{perseguers2010},
 are difficult, at best, to translate to networks of mixed states.
However, these two operations
naturally give rise to a rich set of protocols whose
design is determined by the quantities that are to be optimized.

In Sections~\ref{sec:elements}, \ref{sec:protocols}, and
\ref{sec:analysis}, rather than designing a network for a
particular task, we accept a given network of mixed states
as a constraint. Our goal is then to identify and solve
questions of design that arise in creating protocols to
accomplish entanglement distribution.
%Examples of the protocols are shown in Fig.~\ref{fig:subpaths}b,c.
\begin{figure}
    \begin{center}
        \includegraphics[width=0.85\linewidth]{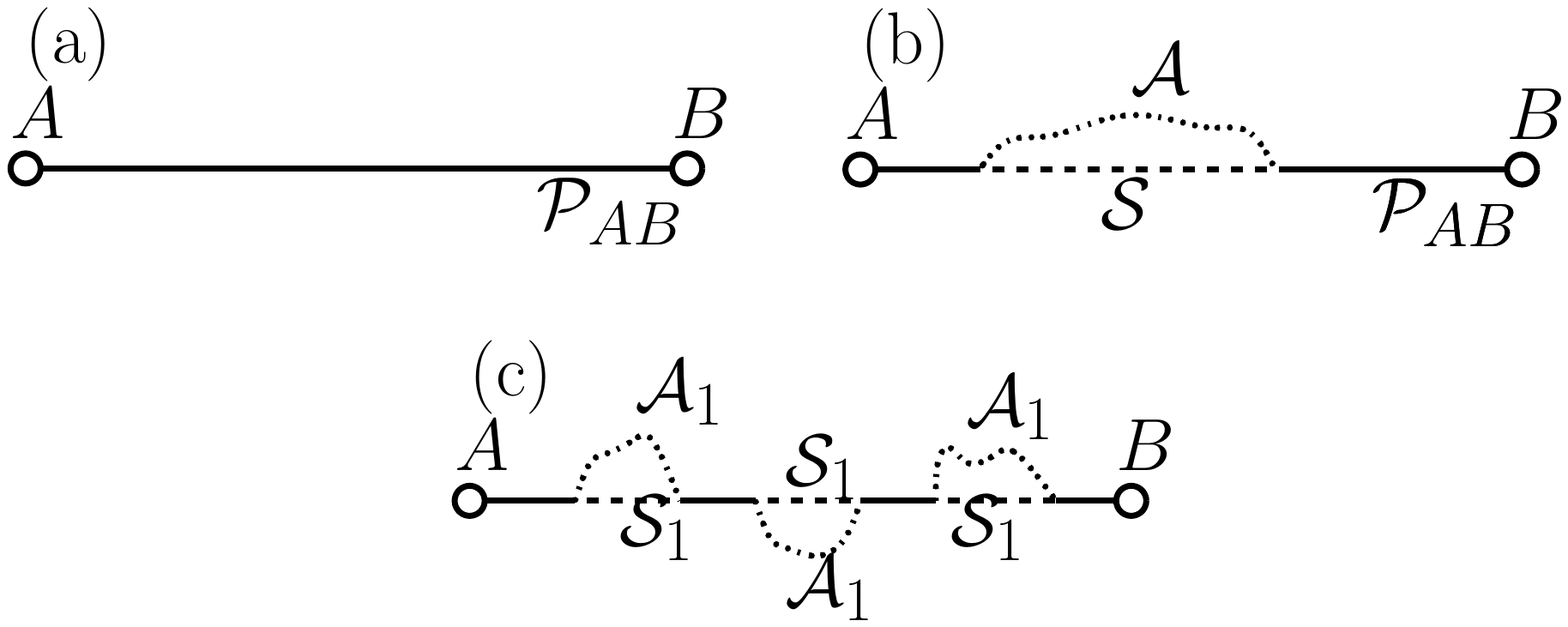}
        \caption{ Establishing entanglement between nodes $A$ and $B$.
  (a) The shortest path $\spath$ between $A$ and $B$; the
geometry of the path is irrelevant, so we represent it by a straight
 line with individual links not shown. Other paths connecting $A$ and
$B$ are not shown.
  (b) The shortest path $\spath$ (solid line with a dashed segment)
  between $A$ and $B$.
Between the endpoints of  subpath ${\cal S}$ (dashed segment) there is an alternate path
${\cal A}$ (dotted line).
 % Note that $\spath$ being the shortest path (or a shortest path)
%implies that $|{\cal S}|\le|{\cal A}|.$
 (c) Similar to (b) with three subpaths and corresponding alternate
paths.}
    \label{fig:subpaths}
    \end{center}
\end{figure}
The main question is: {\it when presented with the option either to swap or to purify,
which is the better choice?} For instance:
\begin{itemize}
\item {\it Single purification protocol. } Consider the
  scenario in Fig.~\ref{fig:subpaths}b in which we want to
  entangle nodes $A$ and $B$ using the shortest connecting
  path $\spath$ while making use of a neighboring path
  ${\cal A}$.  We proceed by swapping at all nodes on the
  two paths ${\cal S}$ and ${\cal A}$ to replace each of
  them by a single link, then purifying these two links,
  followed by performing swappings on all remaining
  nodes. In Sec.~\ref{sec:singpur} we compute the ratio of
  path lengths $|{\cal S}|/|\spath|$ that produces the
  largest average entanglement between $A$ and $B$, finding
  a value of approximately $0.37$.
\item If the goal in the previous example is instead to achieve a positive probability of
entangling $A$ and $B$ with minimal entanglement per link, then the optimal ratio
of path lengths takes the value
$1 + \ln 2 /\ln([\sqrt{5}-1]/4)\approx 0.409$.
\item Consider the scenario shown in Fig.~\ref{fig:subpaths}c,
which we call the {\it multiple purification protocol}.
Here, instead of purifying a single pair of subpaths, we purify $n$ pairs. In Sec~\ref{sec:multpur}
we compute the minimum
entanglement per link required to entangle $A$ and $B$ in the limit of large $n$
for this protocol.
\end{itemize}
In Sec.~\ref{sec:protocols} we define specifically the direct and
quantum strategies mentioned above. In Sec.~\ref{sec:analysis} we
analyze the protocols in a more detailed and quantitative way. In
Sec.~\ref{sec:ernetwork} we apply the single purification protocol
(SPP)  mentioned above to a particular random network--- the
Erd\"os--R\'enyi (ER) network. We present results for short
shortest paths and relatively impure states. We also compute the
exact asymptotic concurrence of the SPP including  all shortest
path lengths at the critical point of the model parameter.
Finally, is Sec.~\ref{sec:noise}, we address the effects on the protocols
of noise in the unitary operations and measurements.

\begin{figure}
    \begin{center}
        \includegraphics[width=.85\linewidth]{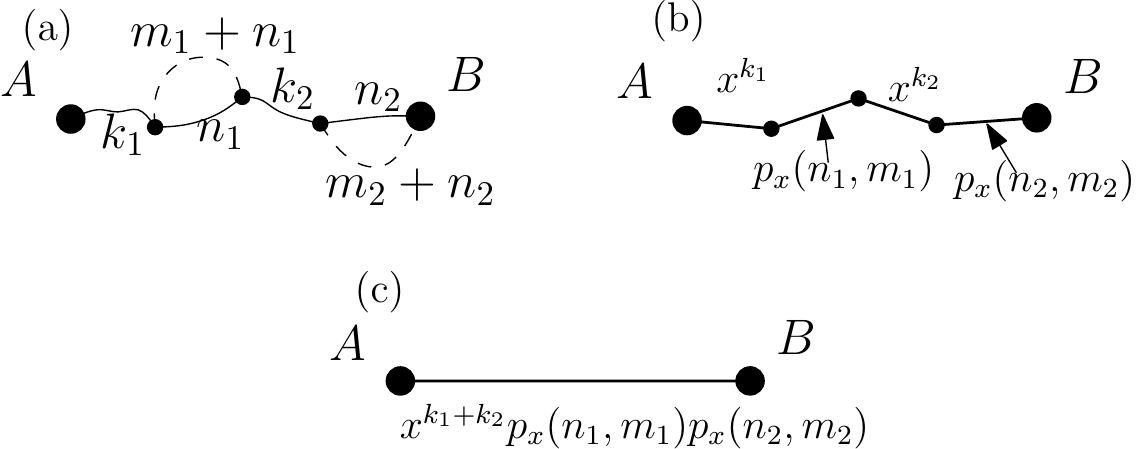}
        \caption{ Quantum protocol. (a),(b), and (c) show the progression of
   a quantum protocol:
   (a) Shortest path between $A$ and $B$ $\spath$ with subpaths of lengths $k_1,k_2,n_1,n_2$. Dotted
  lines show alternate paths of lengths $m_1+n_1$ and  $m_2 + n_2$.
 In this paper, we shall always require that the collection of alternative paths
and $\spath$ be mutually disjoint.
   (b) After swapping and purifying subpaths. Each line segment now
represents a single link with the labels giving the resulting Werner
parameter. (c) After swapping all links.   }
    \label{fig:qprot}
    \end{center}
\end{figure}

\section{Elements of the model \label{sec:elements}}

We first introduce the networks that we shall consider.
We then describe entanglement swapping and purification in
more detail.

\subsection{Network and initial quantum states}

Consider the generic network of nodes and edges shown in Fig~\ref{fig:generic_network}.
With each edge of the network, we associate two
two-level systems forming a bipartite system with states on $\Cf$.
Thus, each node of degree $k$ is
occupied by $k$ qubits.
In the following, we shall consider states diagonal in the Bell
basis
\begin{equation*}
\left\{\ket{\Phi_{ab}}=\frac{1}{\sqrt{2}}\left(\ket{0a}+(-1)^b\ket{1\bar{a}}\right):
 a,b\in\{0,1\} \right\}.
\end{equation*}
%representing states on
%$\mathbb{C}^4$ by their coefficients in this basis.
%In the following, we shall often
%make use of the Bell basis $\{\ketbra{\Phi_{k_1
%    k_2}}{\Phi_{k'_1 k'_2}}\}$,
%$k_1,k_2,k'_1,k'_2\in\{0,1\}$, representing states on
%$\mathbb{C}^4$ by their coefficients in this basis.
In particular, as the initial state on each edge, we choose the Werner
state~\cite{Wer89}
\begin{equation}\label{wernerdef}
 \rho_W(x) = x\ketbra{\Phi_{00}}{\Phi_{00}}+\frac{1-x}{4} \id_4,
\end{equation}
which has fidelity
$F\defeq\bra{\Phi_{00}}\rho_W(x)\ket{\Phi_{00}}=(3x+1)/4$. It can
be shown that the Werner state is that it is entangled for $x>1/3$
and separable otherwise. All protocols in this paper attempt to
entangle two nodes by creating a Werner state on a pair of qubits,
one from each node. Because of its simplicity, the Werner state
serves as a standard form, allowing a clearer exposition of the
distribution protocols than does a Bell-diagonal state. The Werner
state also has the advantage that it is created from any mixed
state by removing the off-diagonal elements via a depolarization
process, a procedure that can be realized by local operations and
classical communication.

\subsection{Concurrence as a measure of entanglement}
In this paper we use concurrence~\cite{Wootters97} as a measure of useful entanglement
in the system. For the Werner state (\ref{wernerdef}) the concurrence is given by
\begin{equation}\label{wernerconc}
C(x)=\max\{0,(3x-1)/2\}.
\end{equation}
Because our task is to entangle any arbitrarily chosen pair of
nodes, we define the average concurrence of the network
\begin{equation}\label{defavcon}
% \avcon{}=\frac{2}{N(N-1)}\sum_{\alpha,\beta}\pi_{\alpha,\beta}C(x_{\alpha,\beta}),
 \avcon{}=\frac{2}{N(N-1)}\sum_{\alpha,\beta}\pi_{\alpha,\beta}C(\alpha,\beta),
\end{equation}
where $x$ is the parameter of the initial state,
$C(\alpha,\beta)=C(x_{\alpha,\beta})$ is the concurrence of the state $x_{\alpha,\beta}$ between
$\alpha$ and $\beta$ after applying some protocol,
and $\pi_{\alpha,\beta}$
is the probability that this protocol succeeds. Thus this definition
depends on the choice of protocol and furthermore assumes that
the concurrence of the resulting state between  $\alpha$ and $\beta$ is zero
with probability $1-\pi_{\alpha,\beta}$. Note that this average is over
both pairs of nodes, as well as probability of the success of
the protocol.

In particular, we judge a particular protocol to be better than
the direct protocol if it yields a higher concurrence averaged
over measurement outcomes.  When applied to our protocols on
two-qubit Werner states, the concurrence has at least two
advantages over other entanglement measures in this respect.
Firstly, the concurrence is the unique entanglement measure that
is linear in $x$, which makes analysis easier. Secondly, the
concurrence provides the extremal comparison in the following
sense. Most of the interesting entanglement measures are either
convex (for instance, entanglement of formation) or concave (for
instance, logarithmic negativity). Suppose that a given protocol
has higher average entanglement than the direct protocol if
concurrence is used as the measure. In Appendix~\ref{sec:appendix}
we show that the protocol also has higher average entanglement, if
any convex entanglement measure is used rather than concurrence.
Conversely, if the protocol is worse than the direct when judged
by concurrence, then it is also worse when judged by any other
concave entanglement measure. Finally, we note that for the state
(\ref{wernerdef}) the concurrence and the negativity are
identical.

\subsection{Operations for distribution and concentration of entanglement}

%\section{Entanglement distribution and concentration}

%\subsection{Entanglement swapping}
%\subsection{Entanglement purification}

% \section{Model network}
% \subsection{Scale-free network model}
% For numerical simulations, we generate a scale-free network according to
% the following algorithm. We first fix parameters $N$ the total number of
% nodes, $k_0$ the smallest degree present, $n_{k_{\text max}}$ the number of nodes
% of largest degree, and $\alpha$.  We then generate a number of nodes for each
% degree $n_k=\lfloor ck^{-\alpha}\rfloor$, where $c$ is determined by the
% condition $\sum_{k=k_0}^{k_{\text max}} n_k = N$.
% for each degree $k$. We then choose randomly a set of edges to connect the
% nodes that is consistent with their chosen degrees $k$. The edges are chosen
% uniformly from all possible such edge sets using the algorithm citealgorithm.

%\section{Entanglement distribution}
%\subsection{Elementary quantum operations}
%In constructing distribution protocols, we restrict the number
%of elementary operations in our palette to two: entanglement
%swapping and entanglement purification. (because we
%don't know of any others?) Together with the topology
%of the network, this already provides a rich array of protocols,
%which we begin to analyze in this paper.

\subsubsection{Entanglement swapping}
In this section, we review entanglement swapping, and
present the result of applying the operation to Werner states.
Consider a state of four qubits $\alpha,\beta,\delta,\gamma$,
such that $(\alpha,\gamma)$ is an entangled pair and $(\delta,\beta)$
is an entangled pair, but systems $\alpha\gamma$ and $\delta\beta$
are in a product state.
Entanglement swapping is a sequence of quantum operations
that transfers entanglement leaving $(\alpha,\beta)$ entangled
and $(\delta,\gamma)$ entangled.
(See Fig.~\ref{fig:swappure}.)
%More precisely, if initially the pair of qubits
%$(\alpha,\gamma)$ are entangled and $(\delta,\beta)$ are
%entangled, then entanglement swapping will result in
%entanglement between $(\alpha,\beta)$ while leaving the
%systems $(\gamma\delta)$ and $(\alpha\beta)$ in a product
%state.
In the case of pure states, the optimal swapping is effected
by measuring $(\gamma,\delta)$ in the appropriate Bell basis, and then
performing a corrective unitary on $\beta$ depending on the
outcome of this measurement~\cite{PhysRevA.60.194}, with the
result being either a maximally, or a partially entangled state
on $(\alpha,\beta)$.  In the latter case, swapping is
usually understood to include an attempted singlet
conversion on $\alpha\beta$, so that the result of the
entire operation is to leave $(\alpha,\beta)$ in either a
maximally entangled state (if successful) or a separable state (if unsuccessful.)
Mathematically, we consider entanglement swapping to be a
map from $\Cf \otimes \Cf$ to $\Cf$, with the reduction of
dimensions resulting from applying a partial trace over the
system $\gamma\delta$. One can show that if the initial states on  $(\alpha,\gamma)$ and
$(\delta,\beta)$ are both maximally entangled ({\it eg} Bell
states), then the swapping operation succeeds with
probability one, assuming perfect operations.  In the case of
Bell-diagonal mixed states, we cannot know which of the four
states we have drawn from the classical ensemble and so
cannot unambiguously interpret the result of a measurement.
The best we are able do is to assume that we have drawn the most
probable state $\ket{\Phi_{00}}$, and proceed with swapping based
on this assumption. However, if we have drawn a state
other than $\ket{\Phi_{00}}$, we are unfortunately
increasing the average classical population
of the remaining Bell states.  More precisely, given two
Bell-diagonal input states whose eigenvalues
are $(A,B,C,D)$ and $(A',B',C',D')$, the un-normalized output
state after swapping is
\begin{equation}\label{swapbelldiag}
\begin{split}
( &AA'+BB'+CC'+DD',\  AB'+BA'+CD'+DC',\\
  &AC'+BD'+CA'+DB',\  AD'+BC'+CB'+DA').\\
%( &AA'+BB'+CC'+DD',\\
%  &AB'+BA'+CD'+DC',\\
%  &AC'+BD'+CA'+DB',\\
%  &AD'+BC'+CB'+DA')\\
\end{split}
\end{equation}
Using (\ref{wernerdef}) and (\ref{swapbelldiag}) it is easy to
compute that performing swapping on two Werner states with
parameters $x$ and $x'$
produces a Werner state with parameter $xx'$. That is,
%compute that performing swapping on two Werner states $\rho_W(x)$ and $\rho_W(x')$
%produces an output state $\rho_W(xx')$.
\begin{equation}\label{swappingformula}
 \rho_W(x)\otimes \rho_W(x') \mapsto \  \rho_W(xx').
\end{equation}

\subsubsection{Purification protocol}  %% Bennett
Purification protocols operate on a collection of bi-partite
mixed states, producing a smaller number of bi-partite states
of higher fidelity than the input states~\cite{dr_entanglement_2007}.
%Purification protocols attempt to increase the entanglement
%in a non-local subsystem by performing  a sequence of local unitaries and measurements
%on the system
We will use the
Bennett-Brassard-Popescu-Schumacher-Smolin-Wootters (BBPSSW)
purification protocol,
 introduced by Bennett {\it et. al.}~\cite{bennett_purification_1996},
which attempts to replace two input Werner states with parameters $x_1$ and $x_2$
by a single, more pure, Werner state--- {\it i.e.} a state with parameter $x'$
satisfying $x'>x_1$ and $x'>x_2$.
The parameter of the state resulting from this protocol is
\begin{equation}\label{wernerpure}
 x'(x_1,x_2) = \frac{x_1+x_2+4x_1 x_2}{3+3 x_1 x_2},
\end{equation}
with  probability
\begin{equation*}
 \frac{1+x_1 x_2}{2},
\end{equation*}
while failure results in two separable ({\it i.e.} useless) states.
Usually, in the study of purification protocols, one is concerned with
the asymptotic limit of repeated purifications. However, in the present
case we are concerned
with a single application of (\ref{wernerpure}). One common situation we
encounter below is purifying two states with $x_1=x_2=x$. Another question
is: Given a state $x$, what is the smallest value of $x_{\text{low}}<x$ such that,
when the states $x$ and $x_{\text{low}}$ are purified, the result is not worse than both of
them; that is $x_{\text{low}}$ for which $x'(x,x_{\text{low}})=x$.
\begin{figure}
    \begin{center}
        \includegraphics[width=0.75\linewidth]{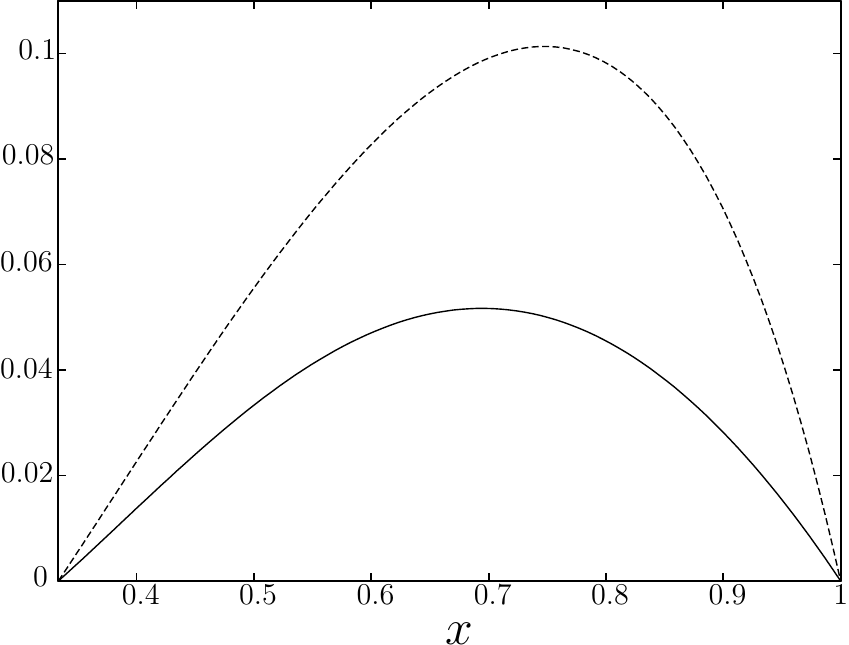}
        \caption{
           $x'(x,x) - x$ (solid curve). $x-x_{\text{low}}$, where
         $x_{\text{low}}$ is determined by  $x'(x,x_{\text{low}})=x$. (dashed curve).
          Both curves cross the $x$-axis at $x=1/3$ and $x=1$.
     }
    \label{fig:purlim}
    \end{center}
\end{figure}
In Fig.~\ref{fig:purlim} we plot $x'(x,x)-x$ and $x-x_{\text{low}}$ {\it v.s.} $x$.
We see that, in order for purification to be useful,  $x_1$  and $x_2$ must not be too
different, and also that purification is most useful for $x\approx 0.7$--$0.8$;
Finally, we note that (\ref{wernerpure}) is increasing in both $x_1$ and $x_2$,
a fact that we will use below.

%From (\ref{wernerpure}) it follows that not all pairs $x_1,x_2$
%produce an output state with $x'$ larger than both the input states.

It is well known that the
Deutsch-Ekert-Jozsa-Macchiavello-Popescu-Sanpera protocol
(DEJMPS), introduced by Deutsch {\it et.
al.}~\cite{deutsch_quantum_1996}, yields states of higher fidelity
than the BBPSSW protocol when performing repeated purifications.
In fact DEJMPS operates on two generic Bell-diagonal states,
producing an output state that is also Bell-diagonal.  When
applied to two Werner states $\rho_W(x_1)$, and $\rho_W(x_2)$ both
protocols yield states with the same fidelity and with the same
probability of success. However, in general, only the output
coefficient of the $\ket{\Phi_{00}}$ component is the same for the
two protocols with the remaining three coefficients differing
between the protocols. In the case that the two input states are a
Werner state and a general Bell-diagonal state with largest
eigenvalue $A$, (\ref{swapbelldiag}) gives a state with
concurrence $[(4A-1)x-1]/2$. This result, together with the fact
that the swapping (\ref{swapbelldiag}) is commutative and
associative, imply that, for protocols using a single
purification, the resulting concurrence is the same whether we use
BBPSSW or DEJMPS. On the other hand, direct calculation shows that
applying (\ref{swapbelldiag}) to two states, each of which is the
result of purifying two Werner states, yields a state whose
concurrence is improved with DEJMPS. In the present work, only the
results in Sec.~\ref{sec:multpur} are non-optimal in this sense.

Finally, we mention that multiparticle recurrence~\cite{PhysRevA.57.R4075},
and hashing~\cite{maneva2002}
protocols have been shown to be more efficient than
protocols operating on two copies.  Improvements have also been made
by optimizing (in part by computer) over a
large class of local unitaries~\cite{PhysRevA.67.022310}
rather than using the unitaries employed in BBPSSW and DEJMPS.
These protocols may give better results, but they are
more opaque conceptually and less amenable to analysis.
Furthermore, the gains shown in other contexts are rather
modest. Thus, we do not consider these more complicated
protocols here.

%One should keep in mind that the formulation of each of these purification protocols
%assumes perfect operations. In practice, the minimum value of $x$ is somewhat
%greater than $1/3$ and the maximum value after repeated purifications
%is somewhat less than $1$.

\section{Entanglement distribution protocols\label{sec:protocols}}

\subsection{Direct strategy }
Our task is to entangle qubits on two selected nodes $A,B$ of the
network. The most naive approach is inspired directly by the
quantum relay: perform repeated entanglement swappings along the
chain of links in the shortest path $\spath$ connecting $A$ and
$B$ using the procedure summarized in (\ref{swappingformula}).
(See Fig.~\ref{fig:subpaths}a.) For instance, swapping the first
two links replaces $\rho_{1,2} = \rho_{2,3} = \rho_W(x)$ with one
new link $\rho_{1,3} = \rho_W(x^2)$.  We then swap the resulting
link $\rho_{1,3}$ with $\rho_{3,4}= \rho_W(x)$, yielding
$\rho_{1,4} = \rho_W(x^3)$, and so on. Thus, after swapping along
the $n$ interior nodes in $\spath$ we obtain $\rho_{A
B}=\rho_W(x^n)$. We call this scheme the direct strategy. It is
somehow analogous to a classical problem of transmission on a
noisy network with transmission probability $x$ on each link.
However, the analogy is not perfect because the classical
transmission probability between $A$ and $B$ $x^n$ is in principle
useful for any $x>0$, whereas in our quantum network any Werner
state with $x<1/3$ is separable and thus useless as a resource for
quantum information tasks. The concurrence of this direct strategy
on a path of $L$ links connecting nodes $\alpha$ and $\beta$ is
\begin{equation*}
 \Ccl(\alpha,\beta) \defeq C(x^L),
\end{equation*}
using $C$ defined in (\ref{wernerconc}).

\subsection{Quantum strategies \label{sec:qms}}
We may improve on the direct approach by using quantum mechanical
operations to concentrate entanglement on the shortest path
$\spath$ connecting $A$ and $B$. In particular, we employ
purification schemes to transfer entanglement from neighboring
paths to subpaths of $\spath$. This follows the general idea of
concentrating entanglement along a ``backbone'' that we used in
previous work~\cite{perseguers:022308}.  But, in the present
setting, we must introduce new techniques because we are not
trying to generate Bell pairs, and we must treat random
neighborhoods of the backbone. In what follows, these more complex
strategies that exploit the network connectivity are called
quantum, although it is clear that the direct protocol is also
quantum.

\subsubsection{Swapping and purifying}
We begin by presenting the elementary combination of the
purification and swapping protocols described above that we
shall use in all of the protocols appearing below.  Consider
two paths, one of $n$ links and the other of $m+n$ links,
with identical Werner states $\rho_W(x)$ on each link, for
instance, paths ${\cal S}$ and ${\cal A}$ in
Fig.~\ref{fig:subpaths}b.  We first perform entanglement
swappings on each chain resulting in two states
$\rho_W(x^n)$ and $\rho_W(x^{m+n})$ which share the nodes at
their endpoints. We then purify these two states to obtain a
Werner state with parameter given by
\begin{equation}\label{bennettpure}
 p_x(n,m) = \frac{x^n+x^{m+n}+4x^{2n+m}}{3 + 3x^{2n+m}},
\end{equation}
the operation succeeding with probability
\begin{equation}\label{probbennettpure}
 \pi_x(n,m) = \frac{1+x^{2n+m}}{2}.
\end{equation}
It is not difficult to prove
 that (\ref{bennettpure}) only yields an improvement over
swapping alone ({\it i.e.} $p_x(n,m)>x_n$) if $m < n$.
%{\bf [not sure how much detail like this to put in appendix or leave out, or what.
% this one can probably be omitted. but it is sometimes difficult to remember what
%we did and did not prove on paper!] }

\subsubsection{Quantum strategy}
Here we present the class of protocols that we study in the
remainder of the paper. In subsequent sections, we will study
particular cases of this class of strategies. These strategies
yield a higher average concurrence than the direct strategy.
Referring to Fig.~\ref{fig:subpaths}a, we say {\bf subpath} for
the segment ${\cal S}$ of $\spath$ that we will purify. We say
{\bf alternate path} for a path ${\cal A}$ disjoint from $\spath$
that we use to purify the subpath  ${\cal S}$.
% This could be explained better and moved to another place.
%Here, pairs of subpaths and alternate paths are only purified once, although in principle, the
%procedure could be iterated.
To entangle a pair of nodes $A$ and $B$, the protocol is as follows.
(See Fig. \ref{fig:qprot}.)
\begin{enumerate}
\item Identify the shortest path $\spath$ between $A$ and $B$ of length $L$. Or, if there
 is more than one shortest path, choose one of them.
\item Identify a subpath ${\cal S}_1$ of $\spath$ with end nodes $a_1$, $b_1$ and length $n_1$,
  such that there is an alternate path ${\cal A}_1$ of length $m_1+n_1$ with $0< m_1 < n_1$ joining
 $a_1$, $b_1$   that is edge-disjoint with $\spath$.
  Note that we cannot have $m_1<0$, because this
 would imply, contrary to our assumption, that $\spath$ is not a shortest path.
\item Repeat step 2 zero or more times, as shown in  Fig.~\ref{fig:subpaths}c,
  finding subpaths of lengths $n_i$ and $m_i+n_i$ edge-disjoint from all previously
 identified paths.  As depicted in Fig. \ref{fig:qprot}(b),
  we now have a collection of subpaths of lengths $n_i$ and $m_i+n_i$ together with subpaths
  of $\spath$ for which there is no sufficiently short alternate path.
\item Perform entanglement swapping at each interior node on each
 of ${\cal S}_i$ and ${\cal A}_i$, effectively replacing each path
  of length $l$ with a single Werner state $\rho_W(x^l)$.
\item Purify each pair of states that resulted from a swapping on
 each pair of paths $({\cal S}_i,{\cal A}_i)$,
 This results in a new path connecting $A$ and $B$
 as shown in Fig. \ref{fig:qprot}(c).
\item Swap along the new path connecting $A$ and $B$ to create a new Werner state between $A$ and $B$
 with parameter
\begin{equation*}
  x'=p_x(n_1,m_1)p_x(n_2,m_2)\cdots x^{L-n_1-n_2-\cdots}.
\end{equation*}
\end{enumerate}

In the following discussion, we find it useful to remove
the length $L$ from all quantities with the following
change of variables.
\begin{equation}\label{varsubst}
y=x^L, \quad a_i=\frac{n_i}{L},\quad b_i=\frac{m_i}{L}.
\end{equation}
Note that $y,a_i,b_i \in [0,1]$, and that $a_i$ and $b_i+a_i$
are now the fractional lengths of the subpath and alternate path,
respectively.
The average concurrence of this quantum protocol is then written
\begin{equation}\label{qmconc}
 \Cqm(\alpha,\beta) \defeq
    \prod_i \pi_y(a_i,b_i) C\left(y^{1-\sum_j a_j}\prod_{i} p_y(a_{i},b_{i})\right),
\end{equation}
where $i$ and $j$ index the purifications in some arbitrary order,
and we have used (\ref{bennettpure}) and (\ref{probbennettpure}).

The choice of subpaths is not specified in the steps above,
but is rather determined by the desired outcome. Below, we
give explicit conditions on the choice of subpaths for
optimizing different quantities: maximum size of interval in
initial fidelity for which QM protocol is better; $n$ that gives minimum
initial fidelity (minimum $x$) for which QM protocol gives positive
concurrence; allowed values $n$ near $L$ for which
QM protocol is better; $n$ that
yields the highest concurrence for fixed $x$.
These protocols can, in principle,
be applied to any network.

% probably superfluous
% For small enough $L$, (perhaps $L\lesssim 15$) we use a computer to
%enumerate all possible combinations of subpaths and order them according to the criteria.
%Below we analyze the constraints for more generic cases.

The basic measure of success of the QM protocol given in steps
$1$--$6$ above is that it must give a better average concurrence
than the direct approach. In the remainder of the paper we shall
often be concerned with the increase in concurrence resulting from
using a QM protocol. We denote this increase between nodes
$\alpha$
 and $\beta$
connected by a shortest path of length $L$ by
\begin{align}\label{deltacdef}
\Delta C(\alpha,\beta) = & \Cqm(\alpha,\beta) - \Ccl(\alpha,\beta)  \\
   =& \Cqm(\alpha,\beta) -   C(y)  \notag.
\end{align}
Likewise $\Delta\bar{C}$ is $\Delta C(\alpha,\beta)$ averaged over a network.
%$\Delta C_{L,n,m}$ is (\ref{deltacdef}) when the path is described by parameters
%$L,n,m$ of a given protocol, etc.
We call the interval in $x$ for which the protocol is successful in this sense
 (plus some reasonable criteria)
the {\bf good interval}. The good interval is determined by the following criteria.
\begin{itemize}
\item  Each pair of subpath and alternative path
 must give an improvement in fidelity after purification. That is,
\begin{equation}\label{constrone}
 p_y(a_i,b_i) > y^{a_i},
\end{equation}
if it succeeds.
This requirement is necessary to avoid protocols which are advantageous,
but would be even better if this particular purification were omitted.
\item  For $y<1/3$, the QM protocol must give a concurrence greater than zero. That is,
\begin{equation}\label{constrtwo}
y^{1-\sum_i a_i}\prod_i p_y(a_i,b_i) > \frac{1}{3}.
\end{equation}
 We call the root of the corresponding equality $\ylo$.
\item  For $y>1/3$, the average concurrence of the quantum protocol
 must be greater than the concurrence of the direct protocol. That is
\begin{equation}\label{constrthree}
\Delta C(\alpha,\beta) > 0.
\end{equation}
We call the root of the corresponding equality $\yhi$.  One
can show that (\ref{constrone}) and (\ref{constrtwo}) give
lower bounds on $y$, while (\ref{constrthree}) gives an
upper bound on $y$.  Physically this can be seen as
follows. If the quantum protocol gives positive concurrence
for some value of $y$, then it will continue to do so for
larger values of $y$ (this also follows from the fact that
(\ref{wernerpure}) is increasing in both arguments). At the
upper bound, the effectiveness of the purification is
decreasing with increasing $y$ (as seen in
Fig.~\ref{fig:purlim}), but the probability of success does
not increase fast enough to make up for the decrease in the
resulting Werner parameter.
\end{itemize}

Finally, we note that the case in which the input parameter
$y<1/3$ (that is $x<(1/3)^{1/L}$) is especially
interesting. In this case, the QM protocol is not only
better on average, but is better in a stronger sense in that
$\Ccl$ vanishes for $y<1/3$.

\section{Analysis of QM protocols \label{sec:analysis}}

\subsection{Generic form of constraints}
In this section, we present the constraints in a form that does not provide
additional insight, but is useful for later calculations.
The constraints (\ref{constrone}),(\ref{constrtwo}), and (\ref{constrthree})
determining the endpoints of the good interval
are each of the form
\begin{equation}\label{yconstreq}
 f(y,\{d_i\}) = \sum_j K_j y^{c_{j,0}+\sum_i c_{j,i} d_i} > 0,
\end{equation}
where $\{d_i\}$ is a relabeling of all of the $a_i$ and $b'_i$,
and $K$ and $c_{j,i}$ are some numbers depending on
$\{d_i\}$ the particular constraint.
We are not interested in the details of this formula, but we use it as a tool for calculating
quantities appearing below.
The end-points of the good interval
are determined by the root, $y^*$ between $0$ and  $1$
of  $f(y^*,\{d_i\})=0$.
The end-point determined by (\ref{yconstreq}) is thus given by
\begin{equation*}
 x^* = {y^*}^\frac{1}{L}.
\end{equation*}
We denote by $\hat y$  the root of
\begin{equation}\label{azeroroot}
 f(\hat{y},\{d_i=0\}) = 0.
\end{equation}
The last expression is useful for computing perturbations around
the solution of equations of constraint that are formulated such that
all parameters vanish.

\subsection{Properties of constraints}

These properties hold for all protocols described by
the six-step procedure above.
\begin{itemize}
\item Consider for the moment just a single subpath of fractional
  length $a$ and the alternate path of fractional length
  $a+b$.  We ask which is better: swapping along the subpath
  and ignoring the alternate path, or swapping along each of
  them and purifying the result. The threshold at which
  purifying yields a Werner parameter equal to the input
  is described by  $\Delta p_y(a,b)=p_y(a,b)-y^a=0$, 
  which defines the
  threshold in each of the parameters  $a,b$, and $y$ implicitly as a function
  of the other two. This equation is easiest to analyze if it
  is reparameterized as $\Delta p_y(a,ca)=0$, that is, by eliminating
  $b$ via $b=ca$. The parameter $c$ is also interesting because it
  gives the fractional excess length of the alternate path relative
  to the subpath. It is not difficult to prove that:

  \noindent {\it i}) The threshold $c$ is given
  by $c=c(y^a)$ where
  \begin{equation}
     c(z) = \ln\left(2[1+4z-3z^2]^{-1}\right)\ln^{-1}(z).
  \end{equation}

  \noindent {\it ii}) $c(z)$ takes valid values ({\it i.e.} non-negative and real)
  only on $z\in[1/3,1]$ where we define $c(1)$ by $\lim_{z\to 1}c(z)=1$.

  \noindent {\it iii}) $c(z)$ increases monotonically in $z=y^a$, so that the
  threshold $c$ increases(decreases) monotonically in $y$ ($a$).

  \noindent {\it iv}) The difference in Werner parameter $\Delta p_y(a,ca)$
  is maximized by $c=0$ for any fixed $z$, but is maximized by non-trivial
  $z$ for fixed $c$. For instance,  $\Delta p_y(a,0)$ maximized over
  $z$ is approximately $0.05$ and is given by a
  root of $3z^4+8z^2-8z+1$ with numerical  value $z\approx 0.69$.

\item Because the map (\ref{varsubst}) from $y^*$ to $x^*$
  is monotonic, the order of end-points of the good interval
  is preserved as $L$
  varies. In fact, the intervals are compressed with
  increasing $L$. Thus, we only need to analyze the rescaled inequalities.

\item The two roots determined by (\ref{constrtwo}) and (\ref{constrthree})
 coincide at $y^*=1/3$. This is because $\Ccl(y)$ vanishes for $y\le 1/3$, and
increases continuously for $y>1/3$. Thus $y=1/3$ is the threshold above which
subtracting $\Ccl(y)$ from $\Cqm$ is necessary to evaluate whether the
QM protocol is useful.
%An example is shown by the crossing of the pairs of curves in Fig.~\ref{fig:yloyhi}.

\item
 It can be proved that the largest absolute increase $\Delta C$ for the QM protocols compared
to the direct, occurs
 at $y=1/3$ for all protocols, that is, the largest $y$ for which
 the direct protocol gives $\Ccl=0$.
 This is shown in Fig.~\ref{fig:boundcurves}.
\end{itemize}
\begin{figure}
    \begin{center}
        \includegraphics[width=0.85\linewidth]{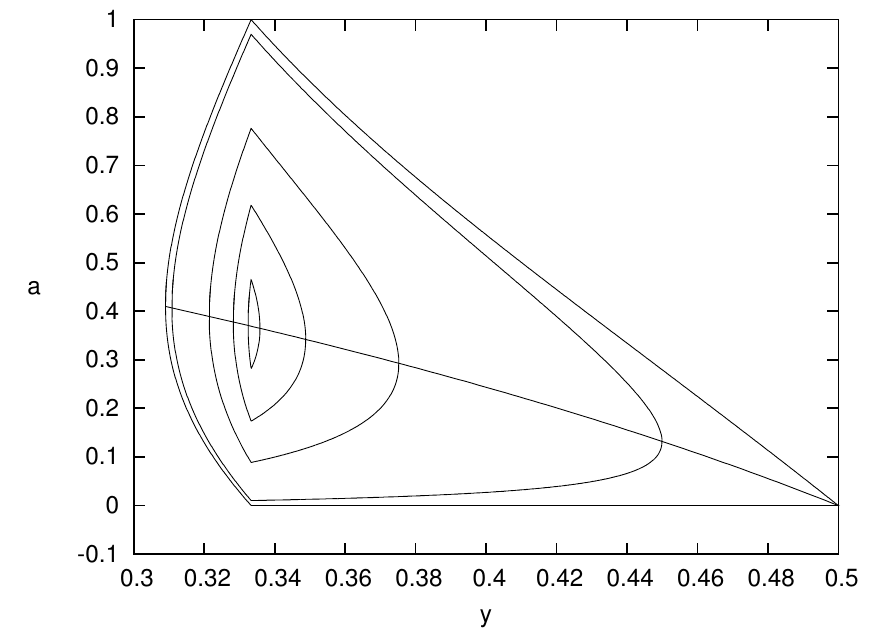}
        \caption{ Roots of $\Delta\Cspp{a,b}(y)=0$ from (\ref{cspp}) as a function of $y$ and $a$ for
          various values of $b$. From the outermost to innermost curve the values of
        $b$ are $0,0.01,0.07,0.11,0.135$.
          Curves are determined from closed-form solutions
          $a=a(y,b)$ with the roots for $b=0$ in particular given by (\ref{hiroots}) and (\ref{loroots}).
          The region inside the closed curves is where $\Delta\Cspp{a,b}(y)>0$ and thus
          SPP is advantageous. The curve cutting through the closed curves is
          $y^a=2y$ and maximizes
          $\Delta\Cspp{a,b}(y)$ with respect to $a$.
      }
    \label{fig:yloyhi}
    \end{center}
\end{figure}
\begin{figure}
    \begin{center}
        \includegraphics[width=0.85\linewidth]{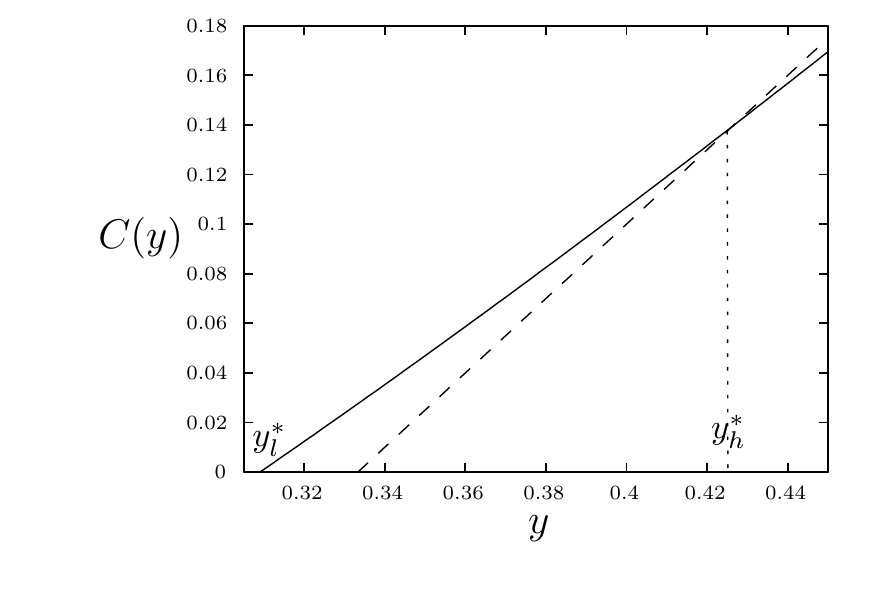}
        \caption{ Average concurrence as a function of
          scaled Werner parameter $y$.  The solid curve is
          $\Cspp{0.409,0}$
          which corresponds to the minimum (at $a=0.409$) of the leftmost
          curve for $b=0$ in Fig.~\ref{fig:yloyhi}. The dashed curve is
          $\Ccl$. Note $\Delta C>0$ for $\ylo<y<\yhi$. The
          lower limit $\ylo$ is determined by the largest
          value of $y$ where $\Cqm=0$.  The upper limit
          $\yhi$ is where the curves coincide.  }
    \label{fig:boundcurves}
    \end{center}
\end{figure}
\begin{figure}
    \begin{center}
        \includegraphics[width=0.8\linewidth]{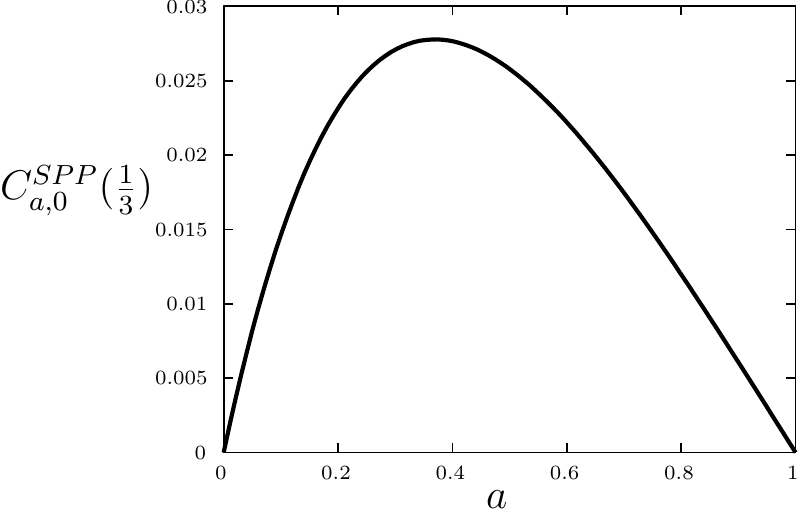}
        \caption{ Concurrence from single purification QM protocol vs. $a=n/L$. For $b=0$ and $y=1/3$.
       This corresponds to the dotted line in Fig. \ref{fig:yloyhi}.
       To produce the maximum concurrence, the optimal fractional size of the subpath is $a\approx 0.369$.
      The curve is generated by setting $b=0$ in (\ref{cspp}), and plotting $C(y=1/3)$ vs.
      $a$.
     }
    \label{fig:conc_vs_a}
    \end{center}
\end{figure}
\begin{figure}
    \begin{center}
        \includegraphics[width=0.85\linewidth]{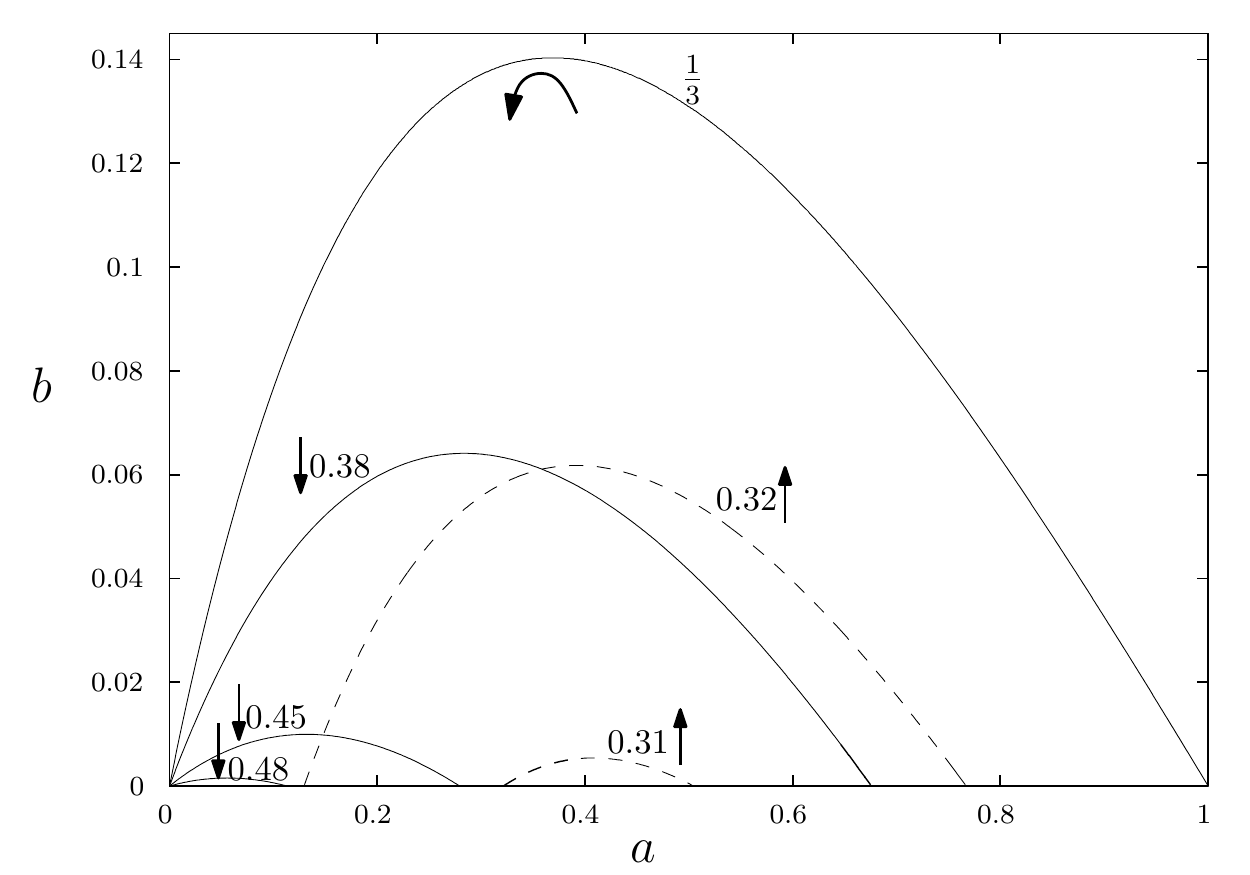}
        \caption{ Region in $ab$-plane for which the single purification QM protocol is better than
      direct ($\Delta\Cspp{a,b}(y)>0$), for several values of $y$. 
      [See (\ref{cspp}).] Curves are labeled with corresponding
      value of $y$. The good area lies below each curve. For $y<1/3$, curves with increasing
  $y$ (dashed lines)  entirely enclose previous areas. For $y>1/3$ the same holds for decreasing $y$. The
   enclosed areas vanish at the smallest $\ylo\approx 0.309$ and the largest $\yhi=1/2$.
   The curves are determined from closed-form solutions $b=b(y,a)$ of the roots of
   (\ref{cspp}).}
    \label{fig:good_area_ab_plane}
    \end{center}
\end{figure}

\subsection{The  single purification protocol \label{sec:singpur}}
We consider here the case of only a single purification
(the single purification protocol [SPP] ) in
which we identify only a single subpath and alternative path
pair. This situation is shown in Fig.~\ref{fig:subpaths}b.
We analyze the protocol finding optimal values according to
the most interesting metrics.
In this case there is only one factor in each of the
products in (\ref{constrtwo}) and (\ref{constrthree}),
while it is easy to see that (\ref{constrone}) is redundant.  Then
(\ref{constrtwo}) and (\ref{constrthree}) become
\begin{equation}\label{tconstrtwo}
 y^{1+b} + 4y^{1+a+b} -y^{2a+b} -g(y) > 0,
\end{equation}
where
\begin{equation}\label{gdef}
 g(y) = \begin{cases}
     1-y &  \text{ for } y< 1/3 \\
     5y-1 & \text{ for } y\ge 1/3.
  \end{cases}
\end{equation}
In accordance with the discussion above, we require
$y,a,b\in(0,1)$.  We call the roots of (\ref{tconstrtwo})
for $y<1/3$ and $y\ge 1/3$, $y^*_{\text{l}}$ and $y^*_{\text{h}}$,
respectively. Explicitly, the increase in concurrence gained from using SPP is
\begin{equation}\label{cspp}
 \Delta\Cspp{a,b}(y) = \frac{1}{4}\left\{  y^b\left[ 4y^2+y-(y^a-2y)^2\right] -g(y) \right\}.
\end{equation}
%\begin{align}\label{cspp}
%\Delta & \Cspp{a,b}(y) = \notag \\
% & \begin{cases}
%   \frac{1}{4} \left(y^{1+b} + 4y^{1+a+b} -y^{2a+b} +y-1\right) & \text{ for } y<1/3 \\
%   \frac{1}{4} \left(y^{1+b} + 4y^{1+a+b} -y^{2a+b} -5y+1\right) & \text{ for } y\ge 1/3
%\end{cases}
%\end{align}
Inspecting (\ref{cspp}), we see that for fixed $y$ and independently of
$b$, $\Delta\Cspp{a,b}(y)$ is maximized for $a$ solving $(y^a-2y)^2$.
If we further maximize over $b$ and $y$, it is not hard to see that
$\Delta\Cspp{a,b}(y)$ assumes a maximum value of $1/36$ at $b=0$, $y=1/3$
and $a=(\log(3)-\log(2))/\log(3)$.
To further illustrate the behavior of (\ref{cspp}), we consider the
simplest case when $a=n/L$, $b=0$, that is, the shortest
path and the alternate path are of the same length. 
The roots of  (\ref{cspp}) solved for $a$ are
\begin{equation}\label{hiroots}
 a(y) = 0, \quad  a(y)=\frac{\log(4y-1)}{\log(y)},
\end{equation}
for $y\ge 1/3$, and
\begin{equation}\label{loroots}
a(y)=\frac{\log\left(2y \pm 2\sqrt{ \left(y-\frac{\sqrt{5}-1}{4}\right)\left(y+\frac{\sqrt{5}+1}{4}\right)}\right)}
          {\log(y)},
\end{equation}
for $y< 1/3$.
In particular, we see that the point where the roots (\ref{loroots})
coincide gives the value of $a$ representing the lowest lower bound
$\ylo$ on $y$ and is given by $a=1 + \ln 2 /\ln([\sqrt{5}-1]/4)\approx 0.409$,
 with  (See Fig.~\ref{fig:yloyhi}.)
\begin{equation}\label{lowestlow}
\ylo=\left(\sqrt{5}-1\right)/4\approx 0.309.
\end{equation}
Thus, this is the optimum value of $a$
to allow the QM protocol to succeed with minimum
initial fidelity. 
On the other hand, the roots (\ref{hiroots}) coincide at the largest
allowed value of $y$,
% As $a\to 0$ the LHS of (\ref{tconstrtwo})
%approaches $0+$, while the root approaches $1/2$, giving a largest
%upper bound at $a=0$ of
\begin{equation}\label{highesthigh}
\yhi= \frac{1}{2}.
\end{equation}
Inspecting (\ref{hiroots}) and (\ref{loroots}) we also see that
the largest good interval in $y$ is obtained for purifying the shortest sub-path,
{\it i.e.} as $a\to0$.
However, the improvement in concurrence also vanishes in this limit.
(See Fig.~\ref{fig:conc_vs_a}). Also note, as shown in Fig.~\ref{fig:conc_vs_a},
that the value of $a$ that maximizes the concurrence is different from the
value that allows minimum initial fidelity as computed above.

We now turn to the case $b\ne 0$ (That is, alternate path is
longer than subpath.) Swapping with a single purification is
in every way worse than if $b=0$. This follows from noting that
the only effect on SPP of increasing $b$ is to introduce a more
weakly entangled state as one of the inputs to the purification.
In particular, there is a value of $a=n/L$ above which the QM
scheme offers no improvement for any value of $y$. One can further show
that the maximum value of $b$ allowing positive $\Delta\Cspp{a,b}(y)$ 
is $b=\log(7/6)/\log(3)\approx 0.14$.
The region in the $ab$-plane for which the single purification protocol yields
an improvement is shown in Fig~\ref{fig:good_area_ab_plane}.

\subsection{Multiple purifications\label{sec:multpur}}
Having analyzed the case in which we are allowed a single
purification, we now turn our attention to the opposite
extreme of unlimited purifications.  We partition a
fraction $\alpha$ of the
shortest path into $n$ subpaths of equal length and purify each
subpath with an alternate path of equal length. We ask how
this protocol performs as $n\to\infty$ and find that the
increase in concurrence tends to a limit, with a lower bound
on $y$ for which the protocol is good given by  $y=(1/3)^{3/(3-\alpha)}$.

Consider $n$ subpaths ${\cal S}_i$ of $\spath$ of lengths
$a_i$, not necessarily covering all of $\spath$, each of
which has a corresponding alternative path ${\cal A}_i$ also
of length $a_i$ (See Fig~\ref{fig:subpaths}c.) We first swap
along each subpath and alternative path, then purify the
resulting pairs. Finally, we swap along the all the
remaining internal nodes.
\begin{figure}
    \begin{center}
        \includegraphics[width=0.85\linewidth]{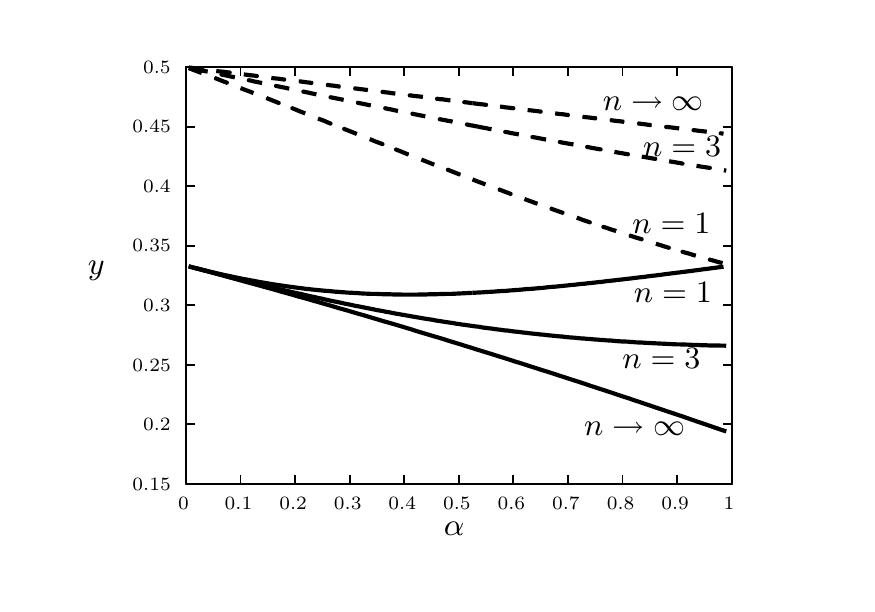}
        \caption{
           Constraints on $y$ for multiple purifications with
$n=1,3,\infty$. The solid curves are $\ylo$.
        The dashed curves are $\yhi$ from the roots of (\ref{multhi}).
$\ylo$ for $n\to\infty$ is taken from (\ref{multlow}), while all
other roots are determined numerically. The values of $y$ between
each pair of solid and dotted lines are the good interval, {\it
ie.} those for which the QM scheme is better  than direct for the
corresponding value of $n$.
     }
    \label{fig:multiple_pur_new}
    \end{center}
\end{figure}
In this case we obtain from (\ref{constrtwo}) the inequality for $\ylo$,
\begin{equation*}
 3\left(\frac{2}{3}\right)^n y \prod_i(1+2y^{a_i})-\prod_i(1+y^{2a_i}) > 0.
\end{equation*}
Likewise the inequality for $\yhi$ obtained from (\ref{constrthree}) is
\begin{equation*}
 \left(\frac{1}{6}\right)^n
 \left[ 2^n 3 y  \prod_i(1+2y^{a_i})-3^n\prod_i(1+y^{2a_i})\right]-3y+1>0
\end{equation*}
In order to investigate the case of purifying many pairs of short paths,
we choose the simplest case, setting $a_i=a$ for all $i$ and $a=\alpha/n$.
That is, we consider purifications
on a fraction $\alpha$ of $\spath$, in which we purify $n$ pairs of paths,
with each path of rescaled length $\alpha/n$.
The inequality for $\ylo$ is then
\begin{equation*}
 3\left(\frac{2}{3}\right)^n y\left(1+2y^\frac{\alpha}{n}\right)^n - \left(1+y^\frac{2\alpha}{n}\right)^n > 0.
\end{equation*}
The limit of the solution of the corresponding equality as $n\to\infty$ is
\begin{equation}\label{multlow}
\ylo=(1/3)^{3/(3-\alpha)}.
\end{equation}
One can show that the inequality for $\yhi$ as  $n\to\infty$ is
\begin{equation}\label{multhi}
3y^{\frac{2\alpha}{3}+1} - y^\alpha -3y +1 < 0,
\end{equation}
which we solve numerically. The results are presented in
Fig.~\ref{fig:multiple_pur_new}, together with the same
curves for a single purification and three purifications.
We saw that the value of $\alpha$ giving the minimum
possible initial entanglement for a single purification is
strictly between $0$ and $1$. However, for two or more
purifications, the minimum is at $\alpha=1$ (this can easily
be proven, as well). Thus, if the goal is for the protocol
to work for the smallest possible initial entanglement, then
performing purifications along the entire path is best in
the present case.  Also, both $\ylo$ and $\yhi$ decrease
with increasing $n$, with the lowest initial entanglement
possible giving non-zero concurrence $y=(1/3)^{3/2}\approx
0.19$ for for $\alpha=1$ and $n\to\infty$. This result
demonstrates that for multiple purifications the best
protocol performs as many purifications on short subpaths as
possible, rather than fewer purifications on longer
subpaths. In this sense, purifying before swapping is
favorable. It seems very likely that the asymptotic limit
mentioned above is the best one can do (with our two allowed
operations) without resorting to using previously purified
links in further purifications.

\subsection{Asymptotic form of constraints}
Here we consider the form of the generic inequalities of
constraint (\ref{yconstreq}) for large $L$, in order to find
simple expressions for the roots, which in turn give the endpoints of
the interval where the quantum protocol is advantageous.
We must take care, however, because we have some choices when taking
this limit.  We consider two different classes of
limits. The first choice is one in which we ignore the
rescaled equations so that $L$ becomes large with $n_i$
fixed. In other words, we are holding the lengths of the
subpaths constant as $L$ becomes large. In this case, we
find that the leading nontrivial term in the root is of
second order in $1/L$.  The other choice is to let $L$
become large with $a_i=n_i/L$ constant. In this case, the
rescaled equations are unchanged in the large $L$ limit so we
only have to look at the asymptotic form of the rescaling
$y=x^L$.  Thus, for large $L$ with $a_i$ held constant, the roots
are given by $ x^* \approx 1 + \ln{y^*}/L$ so that the
interval between two constraints decays as $1/L$. That is,
the length of the good interval in $x$ decreases as $1/L$,
\begin{equation*}
 x_2^*  - x_1^*  \approx \frac{1}{L}(\ln y_2^* - \ln y_1^*).
\end{equation*}
Now we treat the case of holding $n_i$ fixed. We proceed by
first taking the small $a_i$ limit of the rescaled
equations, followed by the large $L$ limit of the inverse
scaling $x=y^{1/L}$.  An expansion of the LHS of
(\ref{yconstreq}) to first order in both $y$ and $a_i$ gives
\begin{equation*}
 y^* = \hat y - \frac{\sum_i a_i \partial_{a_i} f(\hat y,\{{\mathbf 0}\})}
         {\partial_{y} f(\hat y,\{{\mathbf 0}\})},
\end{equation*}
where $\hat y$ is the root of (\ref{azeroroot}).
Replacing $a_i$ by $n_i/L$ and using $(a+b\epsilon)^\epsilon = a^\epsilon + (b/a)\epsilon^2 +O(\epsilon^3)$
we find to second order in $1/L$
\begin{equation}\label{xasymp}
\begin{split}
 x^* = {y^*}^\frac{1}{L} &= {\hat y}^\frac{1}{L}-
   \frac{\sum_i n_i \partial_{a_i} f(\hat y,\{{\mathbf 0}\})}
         {L^2 \hat y \partial_{y} f(\hat y,\{{\mathbf 0}\})} \\
  &= {\hat y}^\frac{1}{L}-
      \frac{ \ln \hat y \sum_{j}\sum_{i=1} K_j n_i c_{j,i} {\hat y}^{c_{j,0}}}
         {L^2 \sum_{j} K_j c_{j,0} {\hat y}^{c_{j,0}}},
\end{split}
\end{equation}
where (\ref{yconstreq}) was used to compute the final line.
%Of course the first term on the RHS includes a term of order $(1/L)^2$. But often,
%constraints are expansions about ${y^*}^{1/L}$, so that, when considering differences
%in the boundaries, the first term does not contribute.
Before proceeding to examples, we make two remarks on the
expansions. {\it i}) for some values
of the parameters $n_i$, the numerator in (\ref{xasymp}) vanishes so the that
the leading term in $1/L$ in the length of the good interval is of order three.
{\it ii}) In some cases, we want to find the limit (\ref{xasymp}) for only a subset
of $\{a_i\}$, with the others held constant. In this case we simply remove some
of the $a_i$ from the sums.

There are several protocols in which these limiting cases are of interest. We mention two of them.
Consider for example a network
disordered in such a way that most shortest paths have nearly the same
length (namely $L$).
That is, if the shortest path between A and B is of length $L$,
then the available subpaths and  alternate paths are
most probably of length near $L$.
Furthermore, we reconsider the scenario in Sec.~\ref{sec:singpur} of a single purification,
 writing $|{\cal S}|=L-q$  and $|{\cal A}|=L-r$
with $L$ large and $q$ and $r$ fixed. Thus $q$ and $r$ represent small
deviations in the length of available subpaths and alternate paths respectively.
In this case, we define $\alpha=q/L$, $\beta=r/L$, and make the substitutions
 $a=1-\alpha$, $b=-\beta$ in (\ref{tconstrtwo}),
and compare the result with (\ref{yconstreq}) taking $\{d_i\}=\{\alpha,\beta\}$
to find the parameters $c_{j,i}$.
Applying (\ref{xasymp}), we find the interval
\begin{equation*}
 \hat x - \frac{\ln 3}{L^2}(3q-2r) < x < \hat x + \frac{\ln 3}{3L^2}(3q-2r).
\end{equation*}
Thus (assuming the roots are analytic in $1/L$)
 only for $q/r > 2/3$, does a a good interval exist for large $L$.

In the single and multiple purification schemes above, we saw
that the optimal length of the alternate path ${\cal A}$
is the same as that of subpath ${\cal S}$, that is $m=0$.
However, an alternate path of length exactly $n$ will
not be available in general.  The lowest order fluctuation
in the upper limit of the good interval, as $m$ varies about $0$,
is studied by examining
the small $b$ limit, with the result
\begin{equation*}
 y^* = \hat y + \frac{m \ln{\hat y} (-\hat y + 4 {\hat y}^{1+a} - {\hat y}^{2a})}
     {L^2(\hat y + 4 (1+a){\hat y}^{1+a} - 2{\hat y}^{2a})}.
\end{equation*}

\section{Application of Single Purification Protocol to Erd\"os--R\'enyi model
  \label{sec:ernetwork}} 
We consider the Erd\"os--R\'enyi (ER) random
graph~\cite{BollobasA,DurrettB} because it is easier to
analyze than more complicated random graphs and gives us
insight into the behavior of the purification protocols on
more complicated graphs.  In particular, we want to compute
the average concurrence under the single purification
protocol (SPP) of section~\ref{sec:singpur} on the ER graph.
The ER model is constructed as follows. Begin with the
complete graph of $N$ nodes and $N(N-1)/2$ edges and then delete
each edge independently with probability $1-p$.  Before
proceeding, we simplify the notation below by introducing
$m'=n+m$ so that the alternate paths are of length $m'$.  In
the following, we call $\sigma_L(p)$ the density of shortest
paths of length $L$ and $\eta_{L,n,m'}(p)$ the density of
SPPs of the given parameters (that is, the fraction of pairs
of nodes that admit this SPP). In general there is more than
one possible position for the subpath of length $n$ along
the SP of length $L$, and $\eta_{L,n,m'}(p)$ includes an
average over these positions.  The most important results in
this section are
\begin{itemize}

\item At low bond densities (small $p$) the density of SPPs characterized
by $L,n,m'$ is proportional to the product of the densities of shortest paths
of length $L$ and length $m'$. That is $\eta_{L,n,m'}\propto \sigma_L \sigma_{m'}$.
The constant of proportionality is determined
by the number of positions for the subpath.

\item At high bond density $\eta_{L,n,m'}\sim \sigma_L$; that is,
most subpaths have an available alternate path.

\item At the critical point $Np=1$, and as $N$ increases,
 all shortest paths are equally likely and the network contains a number
of each possible SPP of order $1$. As $N$ becomes large and
the Werner parameter is near $1$, that is,
 $1-x$ is small, the concurrence gained by
applying all the SPPs is $\Delta\bar{C} \sim AN^{-2} (1-x)^{-4}$ where
$A$ is a constant that is easily  computed numerically.

\end{itemize}

As we saw above, the SPP configurations can be partially
characterized by the numbers $(L,n,m')$ giving the lengths of
the shortest path, the subpath, and the alternate path,
respectively.  In order to compute the average concurrence
for a particular value of Werner parameter $x$, we need to
know the densities for various $L,n,m'$ of the shortest paths
admitting SPP that are beneficial for this value of $x$.  It
would greatly simplify understanding the protocol on complex
networks if we could write the densities of the SPPs in
terms of simpler and better known quantities, such as the
distribution of shortest paths.
  To pursue the connection between these
quantities, we compute below the density of all SPPs on the
ER network in the small $p$ limit and see that in this
limit, the density of an SPP characterized by $L,n,m'$ is
proportional to the product of the density of shortest paths of
length $L$ and the density of shortest paths of length
$m'$. That is, $\eta_{L,n,m'}(p)\approx g(L,n) \sigma_L(p)
\sigma_{m'}(p)$. The factor $g$ is discussed below. On the other hand, we argue that, as $p\to
1$, the density of SPPs with fixed $L,n,m'$ is given simply by
$\eta_{L,n,m'}(p)\approx \sigma_L(p)$.  (That is, nearly all
shortest paths admit SPP). Between these two limits
densities are more difficult to compute. One might expect
similar behavior on other networks that have few connections (small $p$
on the ER network), or many connections (large $p$), but we
have not yet studied other networks in detail.

\subsection{Low Bond Density}
In the limit of low bond density $p$, the numbers
$L,n,m'$ are enough to compute the density of the
corresponding SPP. We take $p$ to be small enough that two
or more SPPs are unlikely to be available for a single pair
of end point nodes $A$ and $B$.
The probability for the SPP configuration is
\begin{equation}\label{etalowp}
\eta_{L,n,m'}(p) = g(L,n) p^{L+m'}\frac{(N-2)!}{(N-L-m')!} +\mathcal{O}(p^{L+m'+1}),
\end{equation}
or $\eta_{L,n,m'}(p)\sim g(L,n)p^{L+m'} N^{L+m'-2}$ for large $N$,
where
\begin{equation}\label{gdegen}
g(L,n)= \begin{cases}
      L-n+1   &  \text{ for } m'\ne  n,\\
      (L-n+1)/2   &  \text{ for } m'= n,\\
\end{cases}
\end{equation}
is computed in Appendix~\ref{sec:compspspp}.
Similarly, we can show
\begin{equation*}
\sigma_L(p) = p^L \frac{(N-2)!}{(N-L-1)!}+\mathcal{O}(p^{L+1}),
\end{equation*}
or $\sigma_L(p)\sim p^{L} N^{L-1}$ for large $N$.
It follows that
$ \eta_{L,n,m'}(p)\sim g(L,n)\sigma_L(p) \sigma_{m'}(p)$
in this limit.

\subsection{High Bond Density}
On the other hand, when $p$ is large enough that a shortest
path of length $L$ is rare, then nearly all shortest paths
are of length less than $L$.
% Since the alternate paths are
%also shortest paths (with different endpoints) of length
%less than $L$, they will be common.
If the SP does not admit an SPP with subpath length $n$, then
an edge-disjoint alternate path must be absent in all $L-n+1$ positions,
which becomes rare with increasing  $p$ and $L$.
It follows that nearly
all shortest paths of length $L$ will allow an SPP for all
possible $n$ and $m'$. Let us consider in particular
$L=3,n=m'=2$.
\begin{figure}
    \begin{center}
        \includegraphics[width=0.85\linewidth]{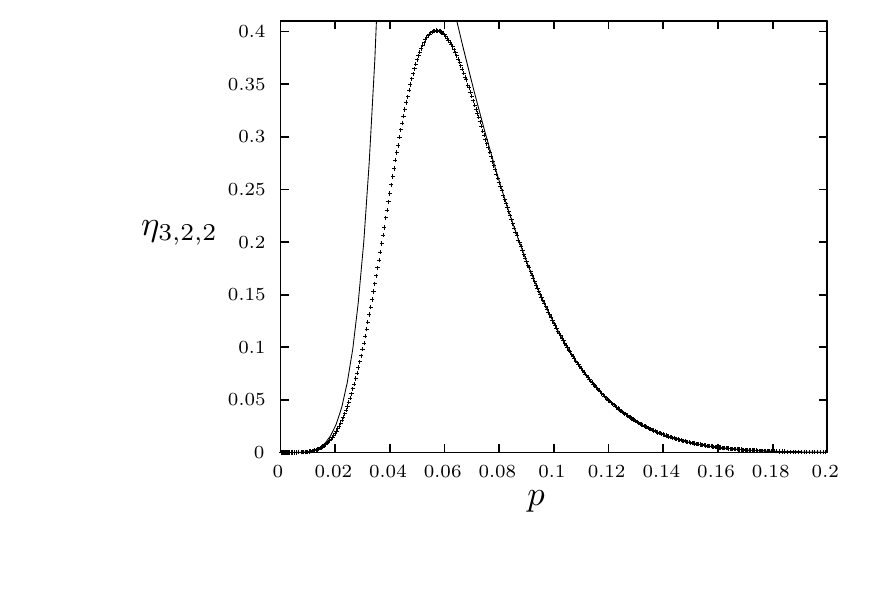}
  \caption{Density of single path purifications $\eta_{3,2,2}$ for
   $L=3,n=2,m'=2$, on  Erd\"os--R\'enyi graph with $N=200$. Points
 are MC data. Curve for smaller $p$ is small the $p$ expansion $p^5(N-2)(N-3)(N-4)$.
 Curve for larger $p$ is the asymptotic formula $(1-p^2)^{N-2}(1-p)$. The
 small and large $p$ regions are shown in more detail using
the same data in Figs.~\ref{fig:eta_approx_lo} and \ref{fig:eta_approx_hi}
 Error bars are not visible on the scale of the plot.
    \label{fig:eta_approx}}
    \end{center}
\end{figure}
\begin{figure}
    \begin{center}
      \includegraphics[width=0.85\linewidth]{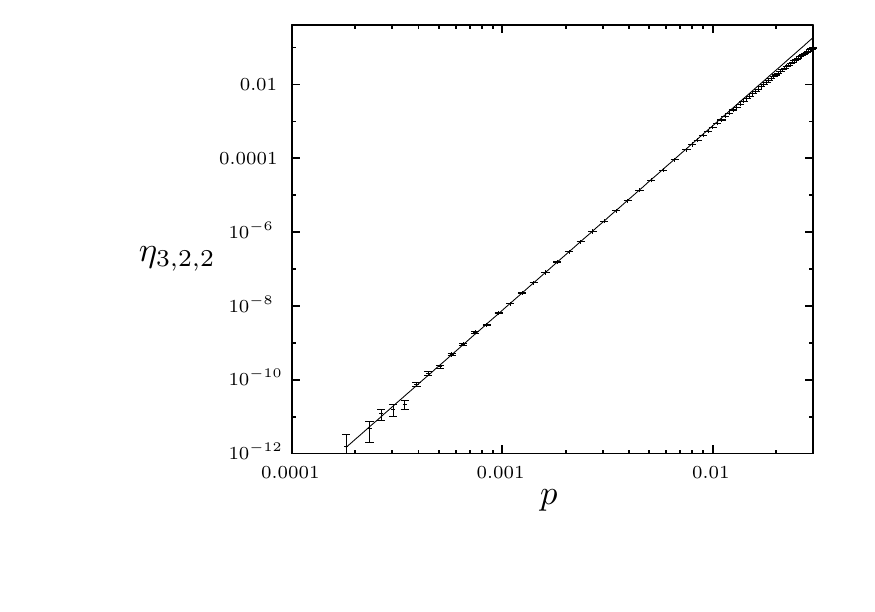}
      \caption{Density of single path purifications $\eta_{3,2,2}$ for
        $L=3,n=2,m'=2$, on  Erd\"os--R\'enyi graph with $N=200$.
        Points are MC data (See Appendix~\ref{sec:MC}.)
          Curve is the small $p$ expansion $p^5(N-2)(N-3)(N-4)$.
    \label{fig:eta_approx_lo}}
    \end{center}
\end{figure}
\begin{figure}
    \begin{center}
      \includegraphics[width=0.85\linewidth]{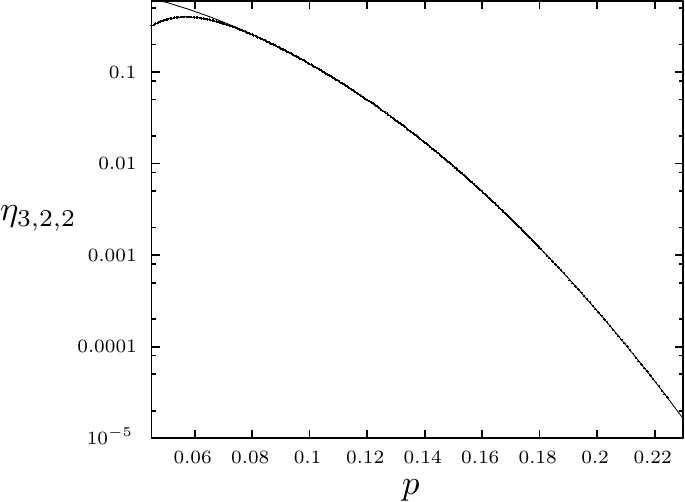}
  \caption{Density of single path purifications $\eta_{3,2,2}$ for
   $L=3,n=2,m'=2$, on  Erd\"os--R\'enyi graph with $N=200$. Points
 are MC data.  Curve is the asymptotic formula $(1-p^2)^{N-2}(1-p)$.
 Error bars are not visible on the scale of the plot.
    \label{fig:eta_approx_hi}}
    \end{center}
\end{figure}
 The density of such SPPs for small $p$ is to lowest order in $p$
$\eta_{3,2,2}(p)= p^5(N-2)(N-3)(N-4)$, as shown in Figs.~\ref{fig:eta_approx}
 and \ref{fig:eta_approx_lo}.  For large $p$, the density of this
SPP is nearly the density of shortest paths of length $3$, which
in turn is nearly the probability that the shortest path is not of length
$1$ or $2$. It is easy to show (see Appendix~\ref{sec:compspspp}) that for all $N$ and $p$,
 $\sigma_1(p) = p$ and
$\sigma_2(p)=(1-(1-p^2)^{N-2})(1-p)$.
We then have $\eta_{3,2,2}\approx 1-\sigma_2 -\sigma_1
 = (1-p^2)^{N-2}(1-p)$ for large enough $p$, as shown in Figs.~\ref{fig:eta_approx}
 and \ref{fig:eta_approx_hi}.
Thus, as we argued in the beginning of this section,
to lowest order in $p$, $\eta_{L,n,m'}\approx \sigma_L \sigma_{m'}$, but this
no longer holds for large $p$ where asymptotically $\eta_{L,n,m'}\approx \sigma_L$.
\begin{figure}
    \begin{center}
      \includegraphics[width=0.8\linewidth]{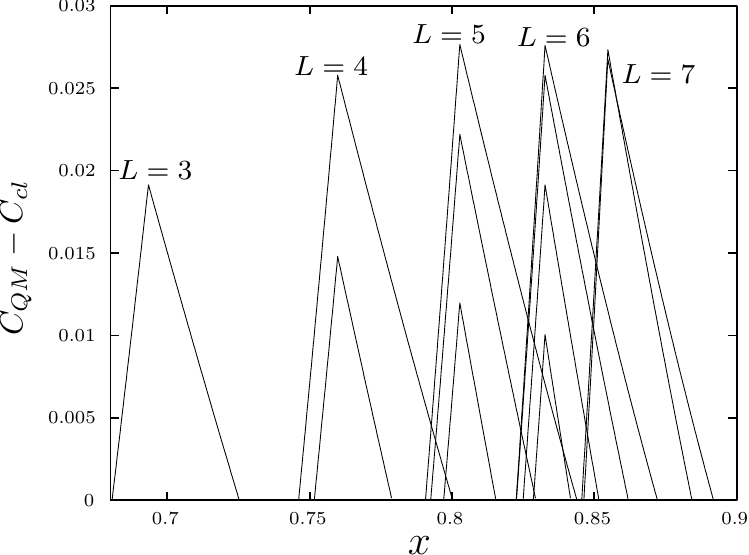}
    \caption{ $\Delta \Cspp{n/L,m'/L}$ (see (\ref{cspp}) given by all possible single
   purifications with $L\le 7$, {\it vs.} the initial Werner parameter $x$. Note that
   this figure is independent of the structure of the lattice. (Some purifications
    for $L=7$ are omitted for clarity.
     }
    \label{fig:cvsx}
    \end{center}
\end{figure}
\begin{figure}
    \begin{center}
        \includegraphics[width=.85\linewidth]{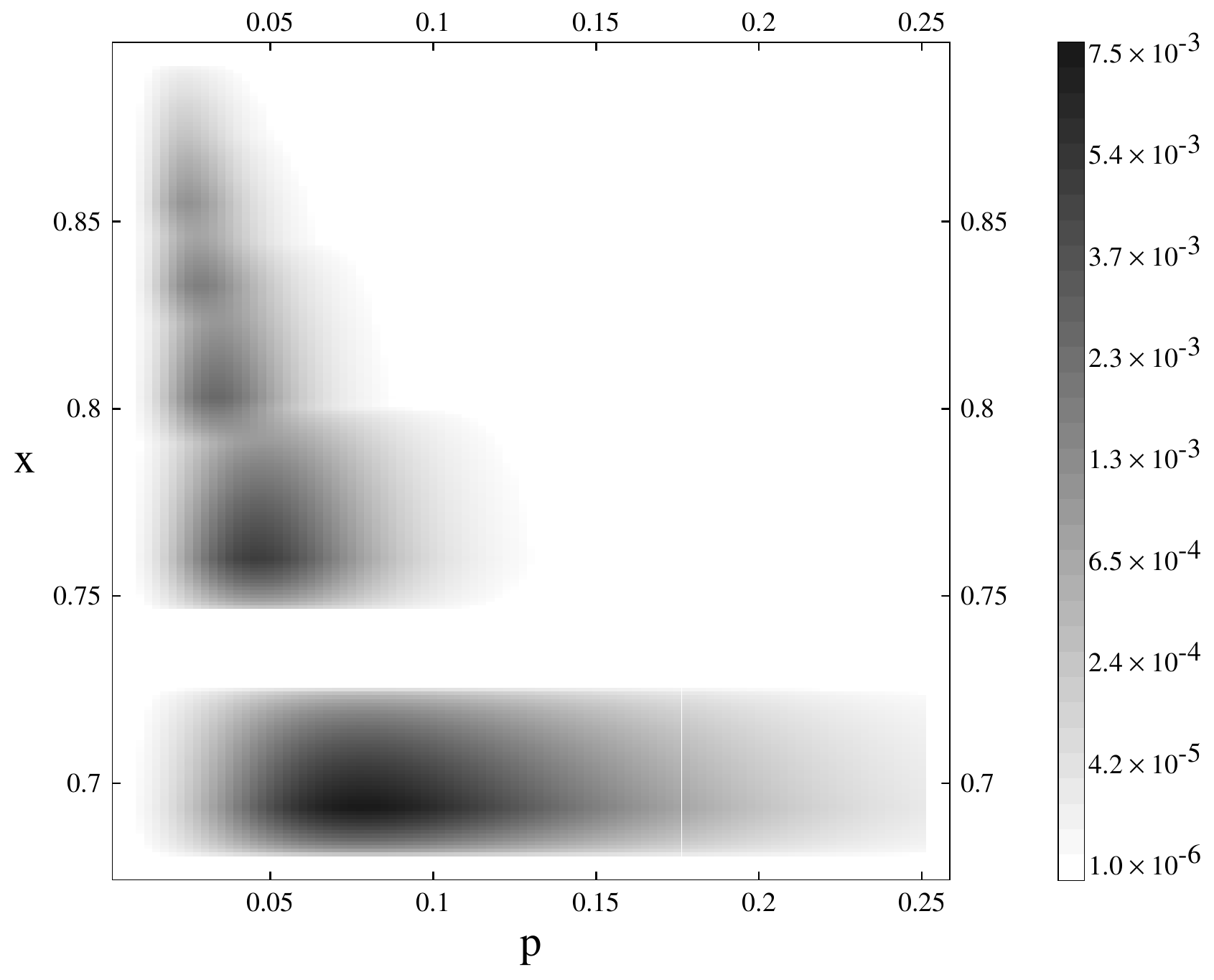}
  \caption{ Gain in concurrence $\Delta\bar{C}$ over the simple method as
  a function of both $p$ and $x$ on the ER network. Monte Carlo results for $N=200$,
  considering all SPPs with $L<8$.
    \label{fig:erdrenCxp}}
    \end{center}
\end{figure}
Finally, we address the consequences of these observations for the
average concurrence. Figure \ref{fig:erdrenCxp} shows the average concurrence as defined in
\ref{defavcon} on an ER network as a function of both input Werner parameter $x$ and
bond density $p$. This plot illustrates several features of the above analysis.
The five concentrations of density correspond to $L=3,4,5,6,7$ (larger $L$
were not computed in the MC calculations). For small $p$, longer SPs and SPPs
are more prevalent, and these require larger $x$ to be effective. On the other
hand, for large enough $p$, most of the SPs are of length $1$ and $2$, which
do not admit SPPs.
\subsection{Concurrence at the critical point $Np=1$}
In this section, we compute the asymptotic average improvement in concurrence
$\Delta\bar{C}$ of the critical ER network for $x$ near $1$ and large $N$ and
find that 
\begin{equation}\label{er_crit_conc}
\Delta\bar{C} \sim \frac{A}{N^{2}(1-x)^{4}}.
\end{equation}
This expression is interpreted as follows. The factor of
$N^{-2}$ is the probability that an SPP with any particular
$L,n,m$ and position of subpath will occur. The factors of
$1/(1-x)$ come from multiple SPPs contributing at one value
of $x$: A fixed value of $x$ gets contributions from SPPs
with associated SPs of length $L\approx 1/(1-x)$; there are
order $L$ such shortest paths; order $L$ different subpaths
(of length $n$) for each SP; order $L$ alternate paths for
each subpath; order $L$ positions along the shortest path
for the subpath.

A few comments on the range of applicability of
(\ref{er_crit_conc}) are in order. In addition to requiring
large $N$ and small $1-x$, we require that contributing
paths not be too large so that the tree-like approximation
remains valid. The most crude bound is that contributing
paths be smaller than the radius (largest geodesic) of the
network. At the critical point, there is a single cluster
of size of order $N^{2/3}$ with all next-largest clusters
growing slower than any power.
  It has been proven recently~\cite{NP08} that the
radius of the incipient giant cluster on the critical ER
graph grows as $N^{1/3}$ and furthermore (in distinction to
the subcritical phase) the smaller clusters have smaller
radii. Our numerical simulations show that the the radius of
the largest cluster is $aN^{1/3}$, with $a$ approximately
equal to $3$. Using this radius as a bound on the valid
range of $L$ together with $L\approx 1/(1-x)$ in
(\ref{er_crit_conc}), we find that $\Delta\bar{C} <
81AN^{-2/3}$. Thus, we see that the advantage
of single-path purification vanishes with increasing $N$
at the critical point of the ER model. We expect similar
behavior on other critical models as they will
also have a broad distribution of very long paths.
On the other hand, if we fix $N=cp^2$, we get asymptotically
$\sigma_2=1-\exp(-c)$ and $\sigma_3=\exp(-c)$, in which case
we expect SPP to continue to show an advantage.

\begin{figure}
    \begin{center}
        \includegraphics[width=.8\linewidth]{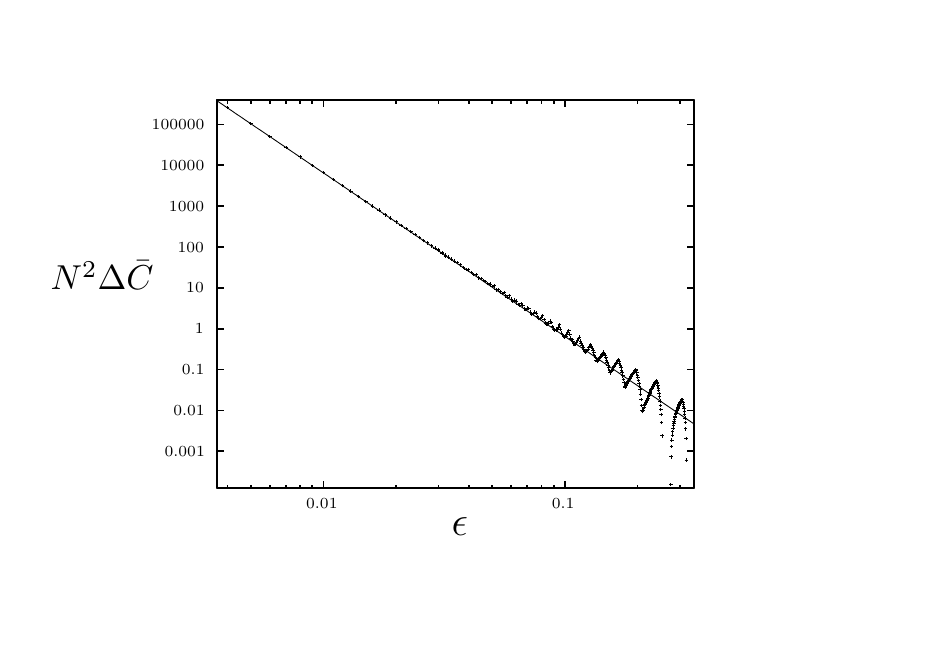}
  \caption{ Average concurrence scaled by $N^2$ {\it v.s}
 $\epsilon=1-x$ on the ER network with $Np=1$.
  The points are computed by summing all contributions of single purifications
 of shortest paths with $L\le 300$. The solid line is the asymptotic result
 $N^2\Delta \bar{C}=A\epsilon^{-4}$. This plot contains no MC, but rather assumes
 $N$ is large enough that each SPP occurs with the probability given by
 the low density expansion (\ref{criteta}). At the rightmost of the plot are the contributions
 from the shortest SPP with $L=3,n=2,m=0$.
     \label{fig:sumL}}
    \end{center}
\end{figure}

The calculation of (\ref{er_crit_conc}) proceeds as follows.  It
follows from (\ref{etalowp}) that
\begin{equation}\label{criteta}
\eta_{L,n,m'}(p=1/N)=g(L,n)p^2=g(L,n)/N^2
\end{equation}
for large $N$.
At this value of $p$ the calculation of the average
concurrence is simplified in that the contributions from
each path admitting SPP have the same dependence on $N$.
Figure \ref{fig:cvsx} shows the contributions to $\Delta\bar{C}$
for individual triples $L,n,m'$, each of which is effective
over a range of $x$. At $Np=1$, we are in the low density regime
and only one SPP is likely to be present between any pair
of vertices. Thus,
for any value of $x$, the total
contribution at $Np=1$ is found by summing over the
contributions for each triple $L,n,m'$.
With increasing $L$, the density of SPPs with nearly the same proportions
(that is, $a$ and $b$) increases. Thus, although all these SPPs are equiprobable,
 as $x$ increases the
contributions come from increasingly large $L$ with the number
of overlapping ranges increasing without limit as $x\to
1$.
In fact, the average concurrence is
\begin{equation}\label{asympconc}
 \Delta\bar{C}(x) \sim \frac{1}{N^2}\int L^3 f(y) \, dL,
\end{equation}
for large $N$ and $x$ near $1$. Here $y=x^L$ and $f(y)$, which
accounts for the sum over $n$ and $m'$, is
computed in Appendix~\ref{sec:critconcurrence}.
We define $h(s)$ via $h(-\ln(y))=f(y)$, and use $-\ln x\approx 1-x=\epsilon$ for
$x$ near $1$. Then integrating (\ref{asympconc}) over $L$ gives
\begin{align}
 \Delta\bar{C}(x) & \sim \frac{1}{N^2} \int L^3 h(\epsilon L)\, dL \notag \\
     & = \frac{1}{N^2 \epsilon^4}
   \int_{s^*_{lo}}^{s^*_{hi}} s^3 h(s) \, ds =  \frac{A}{N^2 \epsilon^4}. \label{gintegral}
\end{align}
The limits on the integral are determined by the lowest lower bound (\ref{lowestlow}) and
largest upper bound (\ref{highesthigh}) on $y$ (as shown in Fig.~\ref{fig:yloyhi}). We
have $s^*_{hi}=-\ln(\yhi)$ and  $s^*_{lo}=-\ln(\ylo)$ with
$A\approx 6.5 \times 10^{-5}$ determined by numeric integration.
In Fig.~\ref{fig:sumL} we see that the asymptotic result (\ref{gintegral}) is approached
rapidly with increasing $L$.

\begin{figure}
    \begin{center}
        \includegraphics[width=.75\linewidth]{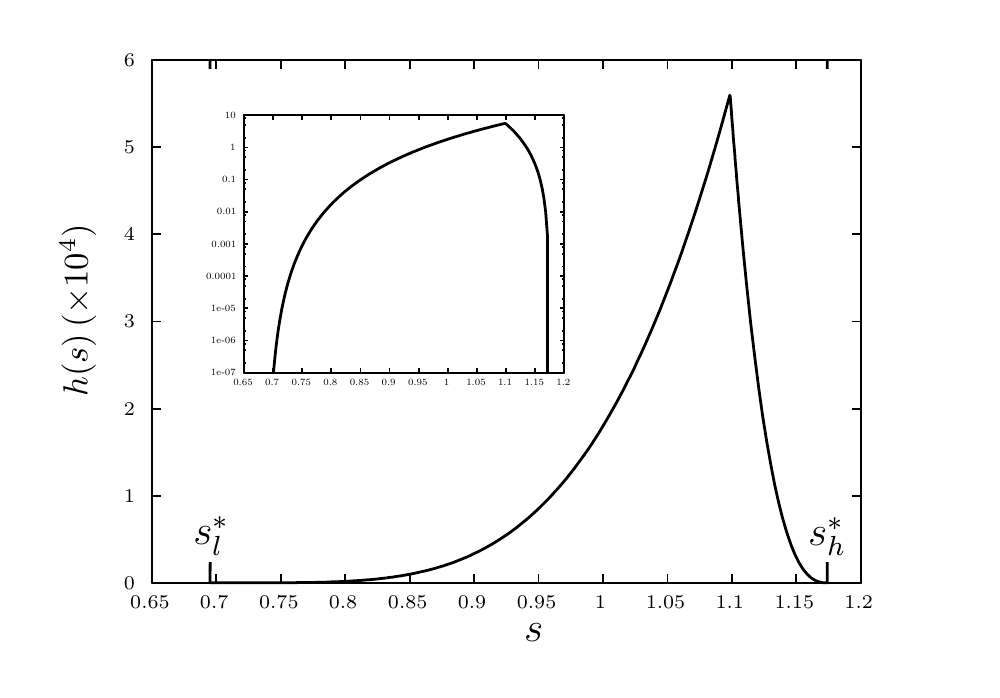}
  \caption{ $h(s)$ appearing in (\ref{gintegral}). Inset is a semi-log plot of $h(s)$
   showing the function vanishing at $s^*_l$  and $s^*_h$ as a power.
  The curve was determined by numeric integration.
    \label{fig:gfunc}}
    \end{center}
\end{figure}

\section{SPP with noisy operations\label{sec:noise}}
Until now, we have considered only the ideal case of perfect operations
and unlimited resources, with the only noise being that inherent in
the Werner state. Even so, the SPP  yields only small improvements in concurrence,
with a maximum improvement of about $0.03$.  However,
because the maximum
concurrence gain from a single successful purifcation using either
BBPSSW or DEJMPS is only $0.05$,  purification 
schemes typically involve repeated purification of a large number of
copies.
Thus, the SPP is conceived as a first step to
investigate possibilities of entanglement concentration on complex networks
and perhaps as a building block in repeated purification schemes.
However, noise can prevent even the small improvement in concurrence from
a single purifcation so that that further purification is not possible.  
In this section we briefly consider the effect of imperfect
unitaries and measurements on the SPP protocol.
We employ a particular noise model for which the effects on
the BBPSSW protocol and the swapping protocol were computed in
reference~\cite{DBCZ99}. Here, a noisy operation is modeled by
a convex combination of the perfect operation and a totally depolarizing
channel that acts only on the same subspace as the perfect operation.
An two-qubit operation on qubits $1$ and $2$ with reliability $p_2$ is described by
\begin{equation}
 O_{12} \rho = p_2 O_{12}^{\text{ideal}} \rho + \frac{1-p_2}{4}\tr_{12}\{\rho\}
         \otimes \id_{12},
\end{equation}
with a similar definition for a single-qubit operator of reliability $p_1$.
An imperfect operation on a single qubit with reliability $p_1$
is described in an analagous way. The imperfect measurement in the
computational basis of a single-qubit
is described by the POVM
\begin{align}
 P_0^\eta &= \eta \ketbra{0}{0} + (1-\eta)\ketbra{1}{1} \\
 P_1^\eta &= \eta \ketbra{1}{1} + (1-\eta)\ketbra{0}{0},
\end{align}
which is a projective measurement only when the parameter $\eta$
is unity.
In~\cite{DBCZ99}, the effects of theses noisy operations
on each the BBPSSW or DEJMPS and swapping protocols was computed.
Although the DEJMPS protocol was reported to be much more robust
against noise, it is also much less amenable to analysis and was
thus treated numerically. Here, we present simple closed-form results
using the BBPSSW protocol. Purification of two input states $\rho_W(x)$
yield a state $\rho_W(x')$ where
\begin{equation}\label{noisybennett}
  x' = \frac{ (2x + 4x^2)(1-\delta) }{3(1+\alpha) + 3x^2(1-2\delta)},
\end{equation}
with $\delta=2\eta(1-\eta)$ and $\alpha=(1-p_2^2)/p_2^2$. The probability
of success is $[1+\alpha + x^2(1-2\delta)]/2$. It is evident that, when $\alpha=\delta=0$,
these reduce to the result for perfect operations given 
by (\ref{wernerpure}).  We assume that swapping with noisy operations a chain of $n$ Werner states each of 
parameter $x$ produces a state
\begin{equation}
 x' = \frac{x^n}{c^{n-1}},
\end{equation}
with parameter $c\ge 1$, which allows us to track
separately the effects of noise from swapping and purification.
If we further assume that the error model for unitaries and measurements described above
applies to swapping as well, then
 reference~\cite{DBCZ99} gives
\begin{equation}
  c = \frac{3}{p_1p_2 (4\eta^2-1)}.
\end{equation}

\begin{figure}
    \begin{center}
        \includegraphics[width=1\linewidth]{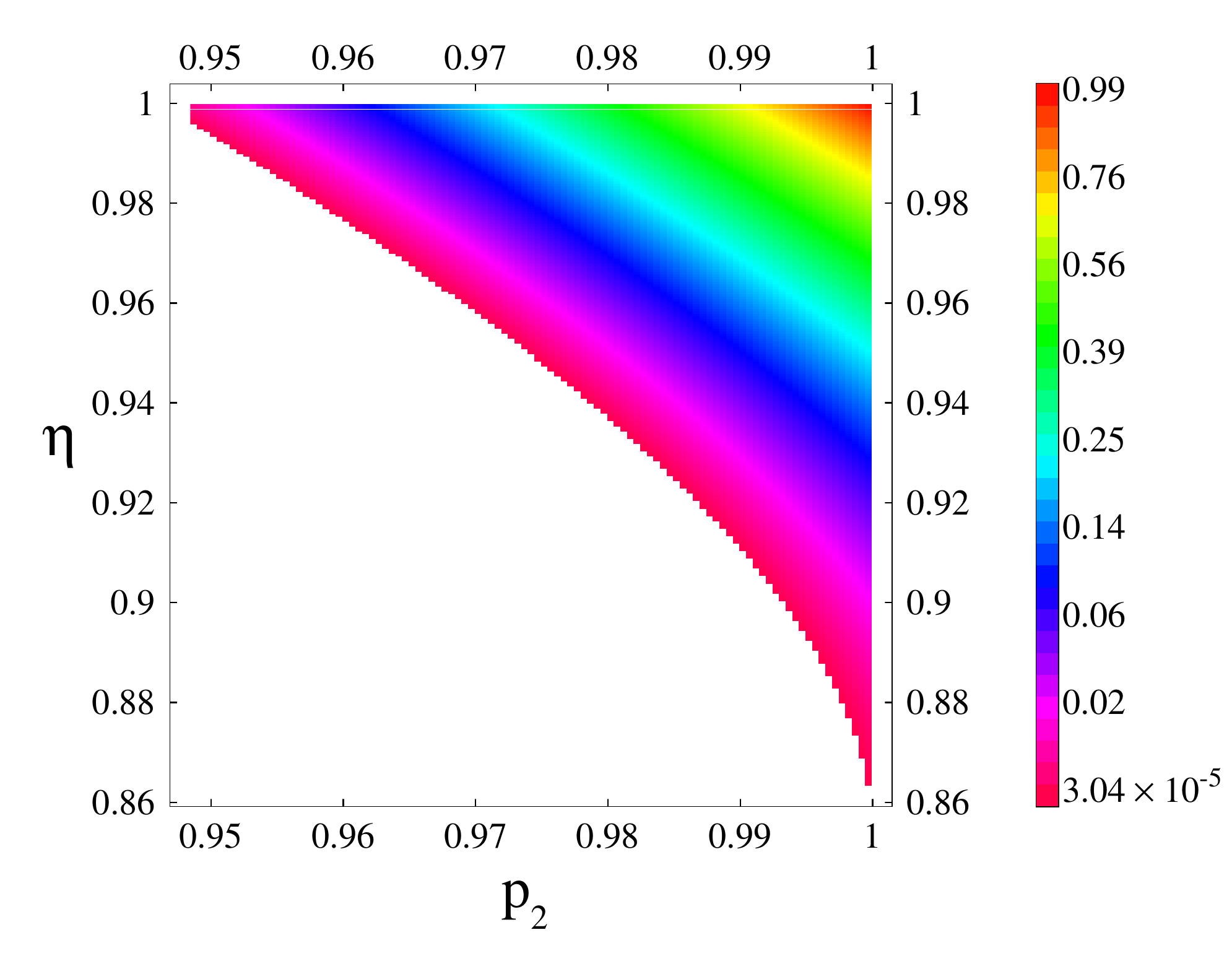}
  \caption{ Maximum value of $\Delta \bar{C}$ with noisy operations,
     as a function of $p_2$ and $\eta$,
      normalized to $1$ for perfect operations.
     \label{fig:cmax}}
    \end{center}
\end{figure}

The rescaled Werner parameter for a geodesic of length $L$ is now
\begin{equation*}
 y = \left(\frac{x}{c}\right)^L.
\end{equation*}
Note that $y$ must now satisfy $y < (1/c)^L$ rather than $y < 1$
as in the case of perfect operations. For simplicity, we restrict
our attention to the case $b=0$ in which the alternate path has optimal length.
The average concurrence (\ref{deltacdef}) with noisy swapping and purification is then
\begin{equation}\label{noisedeltacdef}
\begin{split}
 \Delta  \Cspp{a}(y) = &\frac{1}{4}\Bigg\{ \frac{4c^2(1-\delta)^2}{1-2\delta}y^2 -\tilde g(cy)-2c\delta y  \\
  -\alpha   & - c^2(1-2\delta)\left[y^a-\frac{2(1-\delta)}{1-2\delta}y\right]^2 \Bigg\},
\end{split}
\end{equation}
where
\begin{equation}
\tilde g(w) = \begin{cases}
     1-w & \text{ for } w\le1/3 \\
     4w-1 & \text{ for } w>1/3.
\end{cases}
\end{equation}
To simplify the analysis further, we note that for fixed
$y$, $\Delta \Cspp{a}(y)$ obtains its maximum value at
$a=a_{\text{max}}(y)$ for which the squared expression
containing $y^a$ in (\ref{noisedeltacdef})
vanishes. Furthermore, the maximum over $y$ is obtained for
$y=y_{\text{max}}=1/(3c)$.  This maximum average concurrence
\begin{equation}
 \Delta  \Cspp{\text{max}} = \frac{1}{4} \Bigg\{\frac{4(1-\delta)^2}{9(1-2\delta)}
          -\frac{1}{3}(1+2\delta)-\alpha  \Bigg\},
\end{equation}
which is independent of $c$, is plotted in Fig.~\ref{fig:cmax} {\it v.s.} $p_2$ and $\eta$,
normalized to the value for perfect operations $1/36$. We see that at $y=y_{\text{max}}$ and
$a=a_{\text{max}}(y_{\text{max}})$  for errors of a couple
percent, SSP yields improvements of the same order as for perfect operations. This is
consistent with the sensitivity of the BBPSSW protocol to noise. In fact,
for  $y=y_{\text{max}}$ and $a=a_{\text{max}}(y_{\text{max}})$ the values of the noise parameters
for which $\Delta \Cspp{a}(y)$ vanishes are exactly those for which $x=x'$ in (\ref{noisybennett}).

\section{Conclusion}
We have introduced and solved optimization problems resulting from the interplay
between entanglement distribution and concentration. Already for simple protocols,
the optimal choice of parameters is non-trivial and depends strongly on the
quantity to be optimized.

There are many unexplored questions still to be addressed. For instance, our approach
deals with a static initial network and searches for a protocol with no consideration
of dynamics.

\appendix
\section{Entanglement on two-qubit Werner States}\label{sec:appendix}
In this appendix we show that the concurrence is the extremal
entanglement measure when comparing the quantum to direct
protocols. Below, we show that entanglement measures can be
parameterize by $C$, {\it i.e.} $E=E(C)$. We label two values $E_i
=  E_i(C_i)$ with $i=1,2$. The condition (\ref{qmconc}) has the
form
\begin{equation}\label{qmconcform}
p C_2 \ge C_1,
\end{equation}
with $0<p<1$.
We will show that (\ref{qmconcform}) implies
$p E_2 \ge E_1$ if and only if $E(C)/C$ is non-decreasing, a condition satisfied by
all convex entanglement measures. Thus, if the quantum protocol is advantageous according
to concurrence, then it is advantageous according to all convex entanglement measures.

In this paper, an entanglement measure is a function $E$ from density
operators to $[0,1]$ that satisfies the following conditions

\noindent {\it i}) If $\rho$ is separable then $E(\rho)=0$.

\noindent {\it ii}) $E(\text{Bell state})=1$.

Entanglement measures usually satisfy one or both of two
other properties that will not concern us here: LOCC cannot
increase the expectation value of the entanglement, and for
pure states $E$ reduces to the entropy of entanglement~\cite{PhysRevLett.95.090503}.
Many useful entanglement measures are convex, that is,

\noindent {\it iii)} for positive $p_i$ and $\sum_i p_i=1$,
\begin{equation*}
  \sum_i p_i E(\rho_i) \ge E\left(\sum_i p_i \rho_i\right).
\end{equation*}
Condition {\it i} implies that entanglement measures must
vanish for $x\le 1/3$, so we need not concern ourselves with
these states. The concurrence (\ref{wernerconc}) is an
invertible linear function for states with $x\ge 1/3$, so
they can be parameterize by $C$ rather than $x$, with
$C\in[0,1]$, and we write $E=E(C)$. Because the eigenvalues
of the Werner state (\ref{wernerdef}) are linear in $C$, the
set of states with $x\ge 1/3$ is closed under convex combinations, so that {\it iii}
implies $E(C)$ is convex. Similar statements can be made
about concave functions.
We now show that (\ref{qmconcform}) implies
$p E_2 \ge E_2$ if and only if $E(C)/C$ is non-decreasing.
Clearly, (\ref{qmconcform}) is equivalent to
$p\in[C_1/C_2,1]$. In the worst case, we must then have
$(C_1/C_2) E_2 \ge E_1$, that is, $E(C)/C$ is non-decreasing.
That the converse is true can be shown with similar arguments. Furthermore, it is easy to show that
for all convex entanglement measures $E(C)/C$ is non-decreasing.
A similar argument shows that $pE_2 < E_1$ implies
$pC_2 < C_1$ if and only if $E(C)/C$ is non-increasing,
a condition that is satisfied by all concave entanglement measures.
Finally, it is worth noting that we can make sharper statements. For instance,
$E_a(C) = (C + 4C^2 - C^4)/4$ is neither convex nor concave, yet $E_a(C)/C$
is increasing. It can be shown that the inverse of $E_a$, $E_{a}^{\text{inv}}(C)$,
is also an entanglement measure that is neither convex nor concave, but
 $E_{a}^{\text{inv}}(C)/C$ is decreasing.

%\section{Purifying two paths that have been swapped}

\section{Shortest paths and SPPs on the ER network}\label{sec:compspspp}
In this appendix we compute the density of shortest paths and
paths admitting single purification protocol.

\subsection{shortest paths}

The density of shortest paths of length
$1$ is obviously $\sigma_1(p)=p$. To compute $\sigma_2(p)$,
consider the possible path of length $2$ between vertices $v_a$ and
$v_b$ that passes through $v_c$. This path is absent with
probability $1-p^2$.  Note that the collection of paths for each
of the $N-2$ possible $v_c$ together with the possible path
of length $1$ are mutually
edge-disjoint. Thus the probability that there is a path of
length $2$, but none of length one is $\sigma_2(p) =
(1-(1-p^2)^{N-2})(1-p)$. We did not compute $\sigma_3(p)$,
which would be more difficult because independence is no
longer present. However, we can say something about the case of
large and small $p$. For large $p$, $\sigma(p)_{L+1}/\sigma(p)_{L}$ vanishes
with increasing $p$ so that $\sigma_L(p)\approx 1-\sigma_{L-1}(p)-\sigma_{L-2}(p)\ldots$,
which allows us to compute the asymptotic form of $\sigma_3(p)$. On the other
hand, for small $p$, the probability of more than one path Between $v_a$ and $v_b$
of length $L$ becomes negligible and  $\sigma_L$ is then the sum of the probabilities
for each possible SP. The number of ordered choices of intermediate vertices for the SP
is $(N-2)(N-3)\ldots(N-L-1)$, each of which corresponds to a path with at least one
unique edge. So, for small $p$, $\sigma_L(p)=p^L(N-2)(N-3)\ldots(N-L-1)$. Our Monte
Carlo calculations show that this approximation holds for the case $Np=1$.

\subsection{SPPs}
The computation of density of SPPs is similar to that of
SPs. The number of edges present in an SPP with shortest
path of length $L$, subpath of length $n$, and alternate
path of length $m'$ is $L+m'$. Likewise the number of
intermediate vertices is $L+m'-2$. However, only in the case
$n=m'$, the possible permutations of the $L+m'-2$ vertices
can be partitioned into pairs, in which each member of the
pair defines an SPP including exactly the same edges.
In each pair, the interior vertices of the subpath and
the alternate path are swapped, while the remaining vertices
are unchanged. This
is the origin of the factor of $1/2$ in
(\ref{gdegen}). Finally, the number of ways that the subpath
can be placed along a path of length $L$ is $L-n+1$, from
which we arrive at (\ref{gdegen}).
\begin{figure}
    \begin{center}
        \includegraphics[width=.9\linewidth]{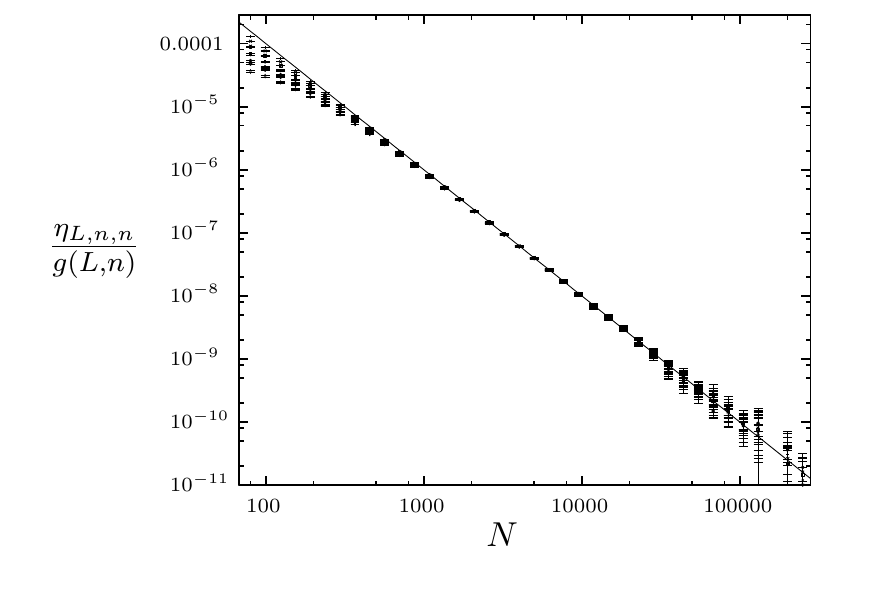}
  \caption{ $\eta_{L,n,n}/g(L,n)$ {\it vs.} $N$ at the critical
  point $Np=1$ for eleven pairs of $L$ and $n$,
  from $3$,$2$, through $7$,$3$.  The solid line is $N^{-2}$ as predicted
 by (\ref{criteta}). Points are MC
  data obtained  by generating ER network samples and counting the number of SPPs.
    \label{fig:etacritMC}}
    \end{center}
\end{figure}
Monte Carlo data supporting this expression is shown in
Fig.~\ref{fig:etacritMC}.

\section{Density of contributions to concurrence at the critical point on the ER network}
\label{sec:critconcurrence}
Here we compute $f(y)$ appearing in (\ref{asympconc}).
In the following we let $p=1/N$.
 We write $\Delta C_{L,n,m}(x)=\max[C^{\text{QM}}_{L,n,m}(x)-C^{\text{Class.}}_{L,n,m}(x),0]$
for the average increase in concurrence obtained from purifying an
SPP with parameters $L,n,m$ between two vertices.
(The average is over quantum outcomes, distribution of ER networks, and possible
positions of the subpath.)
 Here we use $m$  and $m'=n+m$.
As above, we have $b=m/L$ which  takes values between $0$ and $1$.
Then the contribution to the average increase in concurrence
between a pair of vertices for a fixed value of  $L$ is
\begin{align}
 \Delta C_{L}= &\sum_{n=2}^{L-1}\sum_{m=0}^{n-1} \eta_{L,n,m'}(p)\Delta{C}_{L,n,m}(x) \notag \\
  =&\frac{1}{N^2}\sum_{n=2}^{L-1}\sum_{m=0}^{n-1}g(L,n)\Delta{C}_{L,n,m}(x) \notag  \\
  =&\frac{1}{N^2}\sum_{n=0}^{L-1}(L-n+1)\Bigg[\frac{1}{2}\Delta {C}_{L,n,0}(x) \notag  \\
   & + \sum_{m=1}^{n-1}\Delta {C}_{L,n,m}(x)\Bigg] \notag  \\
  =  &\frac{L}{N^2} \sum_{a=2/L}^{1-1/L}(1-a+1/L)\Bigg[\frac{1}{2}\Delta {C}_{1,a,0}(x) \notag  \\
    &+ \sum_{b=1/L}^{a-1/L}\Delta{C}_{1,a,b}(x)\Bigg],  \notag
\end{align}
where we used (\ref{gdegen}), and $a=n/L$, $b=m/L$ take discrete values.
This expression holds for large $N$, but for all $x$ and $L$.
Now for large $L$ we replace the sum with an integral and find
\begin{align}
\Delta C_{L}  =  & \frac{L^2}{N^2}\int_0^1 da \, (1-a)\Bigg[\frac{1}{2}\Delta {C}_{1,a,0}(x)\notag  \\
   &+ L \int_0^1 db \, \Delta {C}_{1,a,b}(x) \Bigg] \notag \\
  = & \frac{L^3}{N^2} \int_{0,0}^{1,1} (1-a)\Delta {C}_{1,a,b}(y)  \, da \, db
        =\frac{L^3}{N^2}f(y).  \notag
\end{align}
In the last line, we discarded the term that is of order $L^2$ and kept the
term of order $L^3$. The final expression is valid for all $x$, but for any
fixed value of $x$, $\Delta C_{L}$ vanishes with increasing $L$ because $y=x^L$.
In practice, we compute $f(y)$ numerically by integrating (\ref{qmconc}) over values $a$ and $b$
for which $\Delta C$ is positive as shown in Fig~\ref{fig:good_area_ab_plane}. The integrand
could be done analytically, but the boundaries in Fig~\ref{fig:good_area_ab_plane} are determined
via numerical roots in any case.

\section{Monte Carlo computations}\label{sec:MC}
We used a modified version of the C language library {\it
  igraph}~\cite{igraph} for Monte Carlo calculations. In
particular, we
% modified shortest path routines and
replaced the calls to the system random number generator with the
Mersenne twister~\cite{mersenne_twister} generator. The
number of trials for computing statistics varies greatly with
model parameters from a few tens to $10^9$. Error bars for
quantities such as the number of SPs and SPPs were computed
from the sample variance in the mean number of events per
network as $\sigma/\sqrt{n_t}$ where $n_t$ is the number of
trials. In a few instances, for the largest error bars,
$n_t$ is not large and the error bars are thus not accurate.

% =============================================================================
% ACKNOWLEDGMENTS, BIBLIOGRAPHY
% =============================================================================
\begin{acknowledgments}
We acknowledge support from ERC grants QUAGATUA and PERCENT, EU
projects AQUTE, Q-Essence, NAMEQUAM, Spanish MINCIN
FIS-2010-14830, FIS-2008-00784, Consolider-Ingenio QOIT projects,
and the Alexander von Humboldt foundation.
% Generalitat de
% the QCCC program of the Elite Network
%of Bavaria, the European QAP and PERCENT projects, the Spanish MEC
%FIS2007-60182 and Consolider-Ingenio QOIT projects, Generalitat de
%Catalunya, Caixa Manresa.
\end{acknowledgments}

\bibliography{qperc}
%\bibliography{main}

\end{document}